\documentclass[pra,showpacs,preprintnumbers,amsmath,amssymb,reprint,twocolumn]{revtex4} 

\usepackage[utf8]{inputenc}

\usepackage{graphicx}
\usepackage{dcolumn}
\usepackage{bm}
\usepackage{hyperref}
\usepackage{color} 
\usepackage{nicefrac}
\usepackage{wasysym}

\usepackage{relsize}

\begin{document}

\title{Control of nonlinear processes using versatile random photonic sources:\\ application to the energy deposition in a dielectric material}

\author{D. Marion\textsuperscript{1}, G. Duchateau\textsuperscript{1,2} and J.-C. Delagnes\textsuperscript{1}}
\affiliation{\textsuperscript{1}CEntre Lasers Intenses et Applications, UMR 5107, Universit\'e de Bordeaux -- CNRS -- CEA, \\ 351 Cours de la Lib\'eration, F-33405 Talence Cedex, France \\}
\affiliation{\textsuperscript{2}CEA-CESTA, 15 avenue des Sabli\`eres, CS60001, F-33116 Le Barp Cedex, France \\}
\email{denis.marion@u-bordeaux.fr}

\date{12 January 2022}

\begin{abstract}
We report on the properties of a non-conventional stochastic photonic source. We first describe the principle of nonlinear intensity filtering by using a modified Mach-Zehnder interferometer. The latter alters the characteristics of an input stochastic source based on Bose-Einstein emission. Computed output intensity fluctuations are compared to an analytical model. Adjusting the interferometer parameters, we show theoretically and numerically that the statistical properties of light such as its probability density function can be tailored. Depending on the parameters, the probability density may exhibit large overshoots or smoother fluctuations. We further evaluate the impact of these modified statistics on simple nonlinear processes. Compared to Bose-Einstein emitters, the yield of nonlinear phenomena varies by several orders of magnitude. We finally simulate the nonlinear interaction of such a laser source with a dielectric material (fused silica) within the framework of a more realistic model, including a statistical analysis. It confirms that the deposited energy varies over many decades and can be largely enhanced due to the properties of the presented laser source.  \textcopyright 2022 American Physical Society 
\end{abstract}

\href{https://doi.org/10.1103/PhysRevA.105.013525}{DOI: 10.1103/PhysRevA.105.013525}
\hspace*{\fill}Please cite as: Phys. Rev. A \textbf{105}, 013525 (2022)


\maketitle

\section{Introduction}
\label{outline:Intro}

Due to their unique features, random photonic sources characterized by partially coherent radiation have  attracted much attention especially for nonlinear light-matter interaction processes. Their propagation in the presence of modulation instability (MI) is analogous to rogue waves in hydrodynamics \citep{suret2016single,akhmediev2016roadmap,Mussot2004}.
Incoherent sources can also be used to measure ultrafast relaxation processes \citep{kobayashi1988application,ando1995large} more easily than in
usual pump-probe schemes. Additionally, optical coherence tomography relies on this kind of sources, that may also be applied in ghost imaging technique \citep{manceauquantum}.

Considering a stochastic light source far from any single or multi-photon resonance, Agarwal \citep{agarwal1970field} established that a $n$-th order nonlinear process evolves as the $n$-th order intensity correlation function. Moreover, it was further established that -- at the same average photon flux -- the rate of nonlinear absorption strongly depends upon the statistics of the source, since its coherence (as defined by Glauber \citep{glauber1963quantum} in 1963) can greatly influence the apparent cross section of multiphoton absorption processes \citep{lambropoulos1967quantum,lambropoulos1968field,PhysRevA.11.1009}.

Compared to a fully coherent laser, an incoherent thermal radiation exhibits a $n!$ (factorial) relative enhancement factor $\eta_n$ in an instantaneous $n$-photon process. Agarwal's result still holds for amplified spontaneous emission (ASE) \citep{Loudon1974}, and for a large class of nonlinear phenomena, including multiphoton ionization \cite{lambropoulos1967multiphoton,PhysRevA.11.1009}, harmonic generation \cite{valero2020} and others.

This phenomenon results from the photon bunching occurring in random \cite{valero2020} or pseudo-random \cite{Muniz-Canovas2019} laser sources. This bunching phenomenon is general to all thermal-like radiations \cite{mandel1959fluctuations}, and sources with a more pronounced bunching effect are referred to as "superbunched" \cite{pegg1986correlations}. Squeezed-vacuum stochastic sources for instance exhibit pronounced enhancements $\eta_n = (2n-1)!! \approx 2^n \cdot n!$  \citep{PhysRevA.36.1288,spasibko2017multiphoton}, that can be further exacerbated by squeezed-vacuum pumped MI \citep{manceauquantum}.

These effects have already been successfully observed and applied in wavelength conversion \citep{valero2020,wu2021tailoring} processes. Experimentally, fluorescence and ASE are two well known examples of incoherent, thermal-like radiations \citep{Loudon1974,saleh2019fundamentals,pietralunga2003photon}. Such sources follow a Bose-Einstein statistics, and the \textcolor{black}{infinitesimal} probability for them to emit $[n,n+dn]$ photons within a given time interval \citep{wong1998photon} is
\begin{equation} 
\label{eq:pGauss}
p_\text{ASE}(\bar{n}, n)\cdot \textcolor{black}{dn} = \frac{dn\cdot\bar{n}^n}{(1+\bar{n})^{1+n}} \approx
\frac{dn}{\bar{n}}\exp({-n/\bar{n}}),
\end{equation}
where $\bar{n}$ is the mean photon number over the same interval. Even for relatively large $n/\bar{n}>2$ values corresponding to multiphoton processes, the probability $p_\text{ASE}$ to observe $n$-photon bunches is then not negligible although located in the tail of the distribution. Let note that the approximation in Eq.~(\ref{eq:pGauss}) only stands if $\bar{n}\gg 1$.

ASE, highly multimode as well as Q-switched sources follow the same thermal-like \citep{Muniz-Canovas2019} or so-called Gaussian distribution. 
Their intensity profile $I(t)$ exhibits random peaks, that can somewhat be considered as the classical equivalent of photon bunching since Eq.(\ref{eq:pGauss}) also reads
\begin{equation}
\label{eq:pGauss_Classical}
p^\text{class}_\text{ASE}(I) ={\left<I\right>}^{-1}\exp({-I/\left<I\right>}),
\end{equation}
where $\left<I\right>\propto\bar{n}$ is the average intensity in the classical limit. Super- and Anti-bunched sources can also be generated by using a cw-seeded frequency-shifted feedback cavity \citep{dechatellus2013generation,PhysRevA.90.033810}. Among the available technologies, active components such as fiber-based Yb${}^{3+}$-doped materials make it possible to conceive broadband ASE sources, powerful enough to observe nonlinear effects \cite{valero2020} and to tackle higher-order ones ($n>2$).

As previously mentioned, the possibility to surpass the $n$-factorial enhancement of ASE involved in a $n$-photon process ($\eta_n = n!$) strongly depends on the coherence and on probability density function (PDF) tail extension \citep{PhysRevA.36.1288,spasibko2017multiphoton,manceauquantum}, but also on the type of nonlinear process. Following these considerations, we theoretically investigate in the present work how a new class of superbunched ($\eta_n > n!$) stochastic sources relying on intensity discrimination based on the nonlinear filtering of an initial ASE source.

Such a superbunched source may result in strongly enhanced higher-order nonlinear effects ($n>2$) by comparing it to the $n!$ enhancement derived by Agarwal in an ideal $n$-th order processes. Since modelling this kind of interactions solely with an instantaneous nonlinearity is a crude approximation, a theoretical proof of concept of this enhancement is also given in the case of intense superbunched sources interacting with dielectric materials, where more realistic interactions are taken into account.

The paper is organized as follows. In section \ref{outline:source description}, we present the concept and properties of a superbunched source based on non-linearly filtered ASE (NLF-ASE). Within an ideal (instantaneous, non-resonant, unsaturated) nonlinear interaction model, we then discuss their enhancement properties, which are later used as input conditions for the laser-dielectric interaction model presented in section \ref{outline:MRE model}. The latter is based on state-of-the-art multiple rate equations (MRE) \cite{Rethfeld} which describe the laser induced nonlinear electron dynamics in dielectrics. The potential of NLF-ASE sources regarding the energy deposition in dielectrics is then provided in section \ref{outline:silica absorption}. The simultaneous presence of the strong isolated peaks on a quasi-continuous background of the NLF-ASE source will be shown as a way to control the laser energy deposition into the material, i.e. the peak-to-background contrast is a key parameter which is provided by the Gaussian PDF for ASE sources. Finally, section \ref{outline:discussion} features important remarks and perspectives regarding the NLF-ASE sources.

\begin{figure}[t!]
\centering
\includegraphics[width=1.\columnwidth]{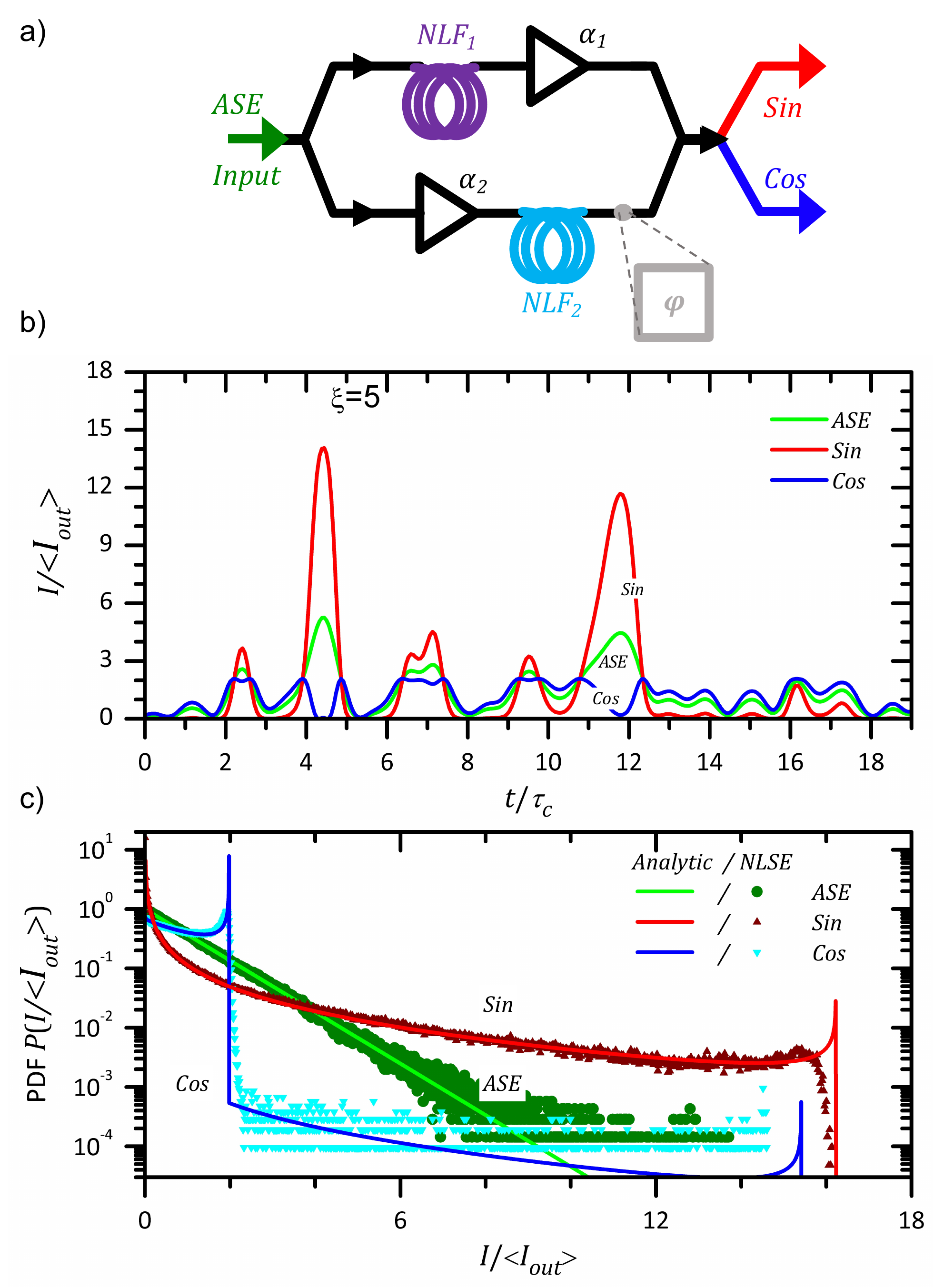}
\caption{Principle of the Nonlinear MZ interferometer (a). Comparison of the relative intensity temporal profiles (b) of the two Sin, Cos NLMZ outputs (normalized to their own average intensity) and the ASE input, and the corresponding probability density functions of signals (c).}
\label{fig:statSources}
\end{figure}

\section{Principle of NLF-ASE sources}
\label{outline:source description}

\subsection{Nonlinear Mach-Zehnder Interferometer}
\label{outline:nlmz}
Prior to the nonlinear filtering through a nonlinear Mach-Zehnder (NLMZ) interferometer (see Fig.~\ref{fig:statSources}(a)), ASE sources exhibit a random intensity profile (see Fig.~\ref{fig:statSources}(b) and Fig.~\ref{fig:AppendixASE}(a) in Appendix~\ref{outline:appendix ASE} for more details) with a characteristic fluctuation timescale given by the coherence time of the emitting atoms. 

The PDF of an ASE source, depicted in Fig.~\ref{fig:statSources}(c), follows an universal exponentially decaying law given in Eq.~(\ref{eq:pGauss_Classical}) regardless of the coherence time. As such, large overshoots naturally take place in an ASE source. They can exceed several times the average intensity $\left<I\right>$: values up to 3 or 6$\cdot\left<I\right>$ are frequently observed, while higher ones are less probable (for instance $\simeq 10^{-5}$ for $I\approx 10\cdot\left<I_\text{ASE}\right>$). 

In the rest of this work, although it cannot be directly measured with conventional techniques, we will assume a short but realistic coherence time (typically 25~fs, corresponding approximately to the Fourier-transform-limited duration of the fluorescence spectrum of Yb${}^{3+}$ in usual lattices) as an initial ASE signal. This is motivated by the fact that, to some extent, stochastic sources can be seen as collection of random short pulses. The more intense and the shorter these pulses are, the larger the nonlinear effects they induce will be.

Nonlinear filtering can actively change the statistical peak-to-background ratio. Among the possible setups such as Sagnac \cite{Sagnac001} or other configurations \cite{NOLM, NALM}, we propose and describe here a NLMZ interferometer that can be easily implemented with current technologies (optical fibre or integrated photonic components \cite{NLZM001, NLMZ002}).

The NLMZ interferometer \citep{shirasaki1990squeezing} schematically depicted in Fig.~\ref{fig:statSources}(a) first consists in a three-ports beam splitter and a four-ports beam combiner. Both the splitting and combining ratios are 50:50. The "upper" branch (1) contains a nonlinear fibre (NLF${}_1$) of length $l$ followed by a gain/attenuation stage ($\alpha_1$). In the "lower" branch (2) the gain/attenuation ($\alpha_2\simeq\alpha_1$) is first applied before propagating through a nonlinear element (NLF${}_2$) of identical length $l$. An additional dephasing element (acousto- or electro-optic modulator, variable delay line) can be added to further manipulate the outputs of the NLMZ, but for the sake of simplicity we restrict ourselves to a zero added dephasing $\varphi$. After recombining, one obtains two outputs referred to as Sin- and Cos-like.

The principle is to introduce a nonlinear dephasing larger in branch (1) than in branch (2). This additional intensity-dependent phase will be interferometrically imprinted on the two outputs, thus modifying the intensity profiles as illustrated in Fig.~\ref{fig:statSources}(b) and the PDFs in Fig.~\ref{fig:statSources}(c).

The characteristics of the gain/attenuation ($\alpha_{1,2}$) and the nonlinear elements NLF${}_{1,2}$ can be finely adjusted so that significantly different nonlinear phases ($B$-integral) are introduced. Additionally their close and low dispersion value will preserve identical time profiles in each branch. Keeping the same intensity profiles in arms (1) and (2) of the NLMZ ensures that the fields on both branches will interfere with the highest possible contrast.

Under such conditions, both Cos and Sin outputs gain very distinctive behaviors, as revealed by closely inspecting Fig.~\ref{fig:statSources}(b) and (c). Note in these figures, each output is normalized by its own average intensity. The Cos output tends to have an homogeneous intensity distribution up to $2\cdot\left<I\right>$, and a near suppression of the stronger peaks.

In contrast, the Sin output has its high-intensity peaks enhanced and all the background of low-intensity peaks is largely attenuated. It also appears that the average time interval between consecutive main peaks lies within the picosecond range.

As discussed later in Section \ref{outline:MRE model}, this value matches the typical characteristic memory time for electron relaxation processes in material contributing to the enhancement of the laser energy deposition along with the intensity. The peak overshoot (Fig.~\ref{fig:statSources}(b), and (c)) is enhanced and can now extend up to $\simeq 16\cdot\left<I\right>$ or more with a much larger probability density than the initial ASE source.

We have checked the feasibility of the NLMZ and computed the two arms intensity profiles $I_1(t)$ and $I_2(t)$ and temporal phases $\varphi_1(t)$ and $\varphi_2(t)$ via nonlinear Schrödinger equation (NLSE) simulations \cite{AgrawalNLFO, fiberdesk} (see Appendix~\ref{outline:appendix PDF NLF-ASE} for details), ensuring that large nonlinear phases ($\sim\pi$) can be obtained with common micro-structured fibers (10~cm long highly nonlinear air/silica photonic crystal fiber, 5~µm mode-field diameter with a nonlinear index $n_2=$ 3.2$\cdot10^{-20}$~m${}^{2}$.W${}^{-1}$) and kW-class (peak intensity) ASE sources \cite{Delagnes}. 

\textcolor{black}{These simulations also confirm that, in the range of spectral bandwidth and peak intensities investigated, only modest spectral broadening is introduced via self-phase modulation (SPM) due to optical Kerr effect.} Owing to the low dispersion values in the fiber considered ($\beta_2=6\cdot10^{-3}$~ps${}^{2}\cdot$m${}^{-1}$), the combined effect of the nonlinearity and the group delay dispersion does not lead to any significant reshaping, thus leaving the temporal intensity profile nearly unchanged.

From the NLSE simulated intensities $I_{1,2}(t)$ and phase $\varphi_{1,2}(t)$, we can compute the two MZ outputs respectively given by
\begin{equation}
    \label{eq: True Sin intensity}
    I_s(t) = \frac{1}{2}\bigg(I_1(t)+I_2(t)-\sqrt{I_1(t) I_2(t)}\cos\left(\Delta\varphi_{12}(t)\right)\bigg)
\end{equation}
for the Sin output, and
\begin{equation}
    \label{eq: True Cos intensity}
    I_c(t) = \frac{1}{2}\bigg(I_1(t)+I_2(t)+\sqrt{I_1(t) I_2(t)}\cos\left(\Delta\varphi_{12}(t)\right)\bigg)
\end{equation}
for the Cos one, with $\Delta\varphi_{12}(t)=\varphi_2(t)-\varphi_1(t)$. 

To allow for a graphical comparison, the PDFs shown in Fig.\ref{fig:statSources}(c) are obtained by computing the histograms of $I_s/\langle I_s\rangle$ and $I_c/\langle I_c\rangle$. Both histograms are then normalized to 1 in integral value. For the case $\xi=5$ chosen here, the transmissions of both arms are approximately $\langle I_s\rangle =1/3\cdot \langle I_\text{ASE}\rangle$ and $\langle I_c\rangle= 2/3\cdot \langle I_\text{ASE}\rangle$. Although it tends to select only moderate to high-intensity peaks, the transmission of the Sin arm is indeed not negligible. These values indicate that both signals could easily be amplified experimentally in fiber amplifiers placed after the NLMZ.

\subsection{Analytical PDF of NLF-ASE}
\label{outline:AnalyticalPDF}

The PDFs of the Sin and Cos outputs of the NLF-ASE source primarily depend on the difference of nonlinear phase $\varphi_\text{NL}=\Delta\varphi_{12}$ accumulated in both arms of the MZ interferometer. As the fiber length $l$ chosen is short enough, the effect of dispersion vs. the Kerr nonlinearity can be neglected. Assuming an identical nonlinear index $n_2$ for both fibers, dispersionless propagation allow us to approximate the nonlinear phase shift by 
\begin{equation}
    \label{eq:phi NL}
    \varphi_\text{NL} = \frac{\pi n_2 l  }{\lambda}\left(1 - \alpha_2\right)\cdot I_\text{ASE}(t),
\end{equation}
where $I_\text{ASE}$ is the input ASE intensity and $\lambda$ is the central wavelength of its spectrum. The additional dephasing $\varphi$ is adjusted so that zero net dephasing (and group delay) is obtained on the Cos port.

Since the shape of the NLF-ASE PDFs obtained from NLSE simulations in Fig.~\ref{fig:statSources}(c) is rather complex, this assumption results in considerable simplifications as the nonlinearity acting on the ASE phase can be treated as instantaneous (local in time) allowing to derive the PDF analytically as discussed below.

If the two arms are balanced ($\alpha_1=\alpha_2$ and 50:50 combiner), the instantaneous intensities of the Sin output may be written as
\begin{equation}
    \label{eq:Sin intensity}
    I_s = \sin^2\left(\frac{\pi}{2}\frac{ I_\text{ASE}(t)}{\xi\cdot\left<I_\text{ASE}\right>_t}\right)\cdot I_\text{ASE}(t),
\end{equation}
while for the Cos output it reads
\begin{equation}
    \label{eq:Cos intensity}
    I_c = \cos^2{\left(\frac{\pi}{2}\frac{ I_\text{ASE}(t)}{\xi\cdot\left<I_\text{ASE}\right>_t}\right)}\cdot I_\text{ASE}(t).
\end{equation}
The critical parameter $\xi$ for an NLF-ASE source is defined as
\begin{equation}
\label{eq:crit parameter def}
    \xi = \frac{\lambda}{n_2 l \left<I_\text{ASE}\right>_t\left(1-\alpha_2\right)}, 
\end{equation}
where $\left<\cdot\right>_t$ denotes the time average.

The PDF $p_{s,\xi}$ of a Sin-like NLF-ASE source of critical parameter $\xi$ may be derived as the series
\begin{equation}
\label{eq:p_s alpha}
p_{s,\xi}(I_s) =  \frac{1}{\left<I_\text{ASE}\right>}\mathlarger{\sum}_{k=0}^{+\infty}\frac{e ^{-\frac{i_k}{\left<I_\text{ASE}\right>}}}{\displaystyle \left|\sin^2{\theta_k}+2\theta_k\sin{\theta_k}\cos{\theta_k}\right|},
\end{equation}
where the $(i_k)_{k}$ are the antecedents of $I_s = f_s(I_\text{ASE})$ (see Eq.~(\ref{eq:Sin intensity})) and $\theta_k = \pi i_k/2 \xi\langle I_\text{ASE}\rangle_t$. Identically, the PDF for the Cos port, $p_{c,\xi}$ can be found in Appendix~\ref{outline:appendix PDF NLF-ASE} along with a full derivation of $p_{s,\xi}$, $p_{c,\xi}$ and a full definition of the $(i_k)_k$ set. Eq.~(\ref{eq:p_s alpha}) allows us to predict the exact shape of the PDF of an experimental NLF-ASE source. As explained before, in the following we only discuss the results for Sin- and Cos-like situations where no extra linear phase $\varphi$ is added. However, it should be mentioned that varying $\varphi$ leads to a gradual change in the PDF shape from the Sin case to the Cos-output one, making the NLMZ a versatile solution.

NLF-ASE PDFs exhibit a step-like structure (Fig.~\ref{fig:statSources}(c), \ref{fig:sim_temp_adv}(b) and \ref{fig:p_s_adv}) that drastically departs from the initial exponentially decaying ASE statistics (Fig.~\ref{fig:statSources}(c)). Peaks of well-defined heights are more often observed (see Appendix~\ref{outline:appendix PDF NLF-ASE} Fig.~\ref{fig:sim_temp_adv}). Their positions and heights depend on the value of $\xi$ and are given by the poles of Eq.~(\ref{eq:p_s alpha}). Here the parameter $\xi=5$ has been chosen so that the PDF for the Sin output, visible in  Fig.~\ref{fig:statSources}(c), exhibits a single pole at $I\approx 16\cdot\left<I_s\right>$ with a significant probability. The corresponding Cos output has a main pole at $I\approx 2\cdot\left<I_c\right>$.

The agreement between the analytical PDFs and those computed from NLSE simulations is excellent, as shown in Fig.~\ref{fig:statSources}(c). The slight discrepancy located in the vicinity of the steps associated with the poles is due to the interplay between the Kerr nonlinearity and the dispersion that has been neglected in the analytical derivation above. This tends to spread the peak in time thus smearing out the abrupt transition around each pole responsible for the peak selection.

The nonlinear peak selection process can be understood as follows: for a given $\xi\gg 1$ associated with a low MZ non-linearity, only the highest peaks which intensity $I_\text{ASE}(t)\gtrsim\xi\cdot\left<I_\text{ASE}\right>_t$ will be efficiently transmitted through the NLMZ Sin-output as suggested by the notation in Eq.~(\ref{eq:Sin intensity}). Conversely, the Cos-output will better transmit low intensity values, thus suppressing the peaks and leading to an "intensity-compressed" distribution. 

Adjusting $\varphi$ and $\xi$ hence allows to control the energy ratio between the low-intensity, continuous-background peaks, and the contribution of sparsely-occurring high-intensity peaks. As discussed in Section \ref{outline:silica absorption}, this feature may be interesting notably for non-instantaneous nonlinear phenomena with a specific characteristic time, such as light absorption in dielectrics where both cumulative and instantaneous effects are involved in the laser-induced electron dynamics.


\subsection{Enhancement factor for an ideal $n$-photon interaction process}
\label{outline:nlMZ simulation}

As discussed in Section \ref{outline:Intro}, the rate of $n$-th order instantaneous nonlinear processes evolves according to $g^{(n)}(0)$ \citep{agarwal1970field}, with
\begin{equation}
\label{eq:Def_g_n_0}
g^{(n)}(0)=\frac{\left<I^{n}(t)\right>_t}{\left<I(t)\right>_t^n} .
\end{equation}
The evolution of $g^{(n)}(0)$ (see Appendix \ref{outline:g2gn} for $g^{(n)}(\tau)$) upon nonlinear filtering is thus of prime importance in order to evaluate the influence of NLF-ASE on multiphoton processes and other highly nonlinear phenomena.

\begin{figure}[t!]
\centering
\includegraphics[width=\columnwidth]{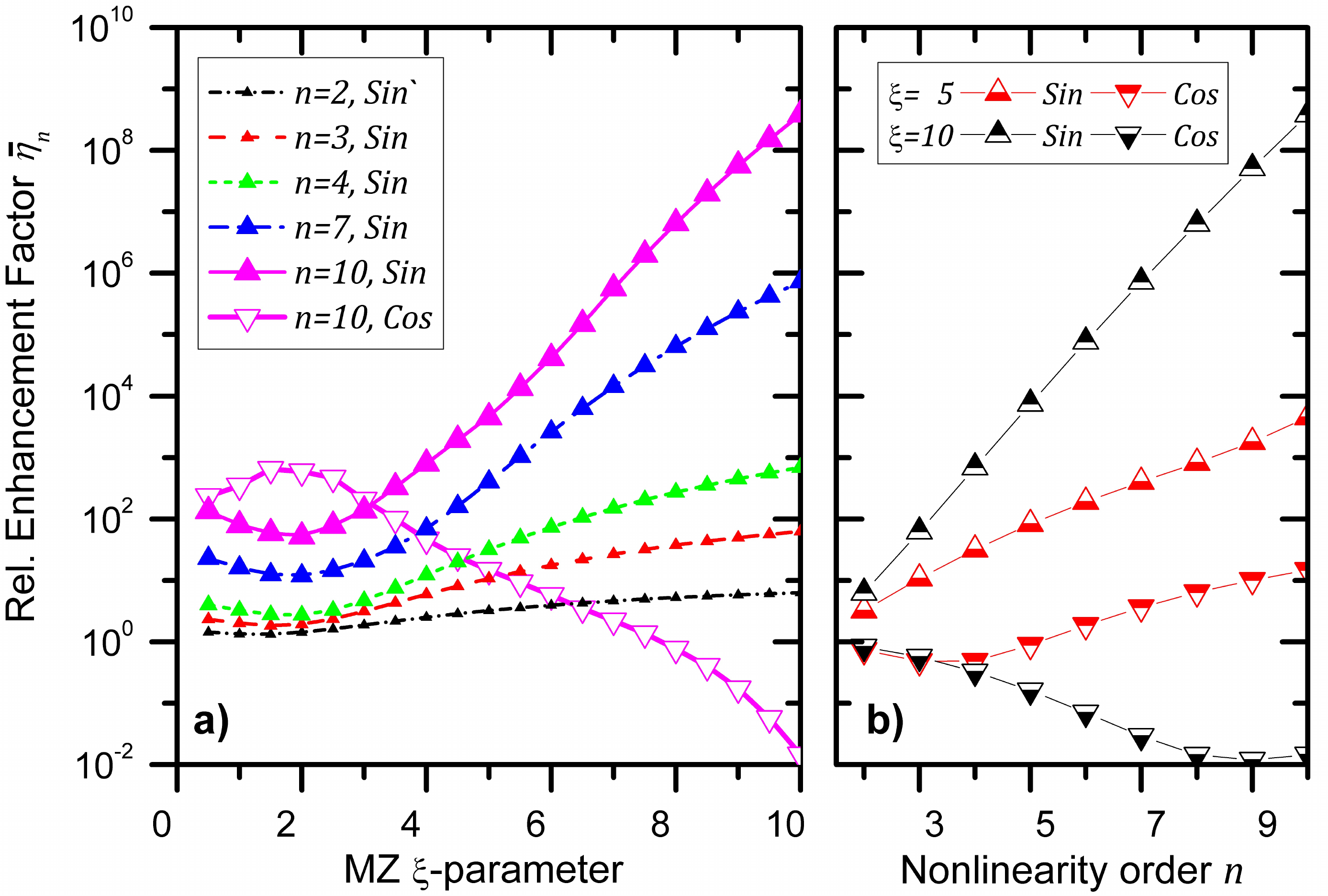}
\caption{Relative enhancement factor $\bar{\eta}_n$ between ASE and NLF-ASE in $n$-th order processes: (a) as a function of the critical nonlinear Mach-Zehnder parameter $\xi$. Upward (resp. downward) closed (resp. open) triangle symbols correspond to Sin-output (resp. Cos). Each curve is associated with a $n=$2, 3, 4, 7, and 10-photon process for the Sin-ouput (resp. n=10 Cos-output). (b) as a function of the order $n$. Upward (resp. downward) triangle symbols $\vartriangle$ ($\triangledown$) correspond to Sin-output (resp. Cos). Each curve set corresponds to $\xi=5$ (red) and 10 (black).
\label{fig:etanfactorial_adv}
}
\end{figure}

It is well known that for incoherent light following the Bose-Einstein distribution, $g^{(2)}_\text{inc}(0)=2$ \cite{Loudon1974,dechatellus2009coherence}, and more generally $g^{(n)}_\text{inc}(0)=n!$. This may serve as a relative point of comparison. The relative enhancement factor
\begin{equation}
\label{eq:G_m_relative}
\eta_n=\frac{g^{(n)}(0)}{g^{(n)}_\text{inc}(0)} = \frac{\left<I^n\right>_t}{n!\left<I\right>_t^n}
\end{equation}
is difficult to derive analytically in the case of nonlinear filtering, but can be readily computed. The input  $I_\text{ASE}(t)$ is simulated using the method described in \cite{valero2020} and  recalled in Appendix~\ref{outline:appendix ASE}. The coherence time of the electric field was set to $\tau_c = 25$~fs and the time increment to $1$~fs. The average enhancement $\bar{\eta}_n=\left<\eta_n\right>_{N_p}$ obtained over $N_p=10000$ independent random realizations of $I_\text{ASE}(t)$ (determining $I_s(t)$ and $I_c(t)$) of 25~ps each vs. the critical nonlinear filtering parameter $\xi$ for different $n$-th order processes is shown in Fig.~\ref{fig:etanfactorial_adv}.

Above $\xi=2$, the relative enhancement $\bar{\eta}_n$ of the Sin output (upward triangles Fig.\ref{fig:etanfactorial_adv}) monotonically increases with $\xi$ (Fig.~\ref{fig:etanfactorial_adv}(a)), while the Cos output (downward triangles) decreases. Large $\xi$ values correspond to low nonlinear MZ dephasing $\varphi_\text{NL}$: within an isolated ASE peak (see Figs.~\ref{fig:statSources}(b)) nonlinear filtering results in high (resp. low) peak-to-background value in the Sin (resp. Cos) output. Nonlinear phenomena will thus be stronger for the Sin- than for the Cos-output. Since $g^{(n)}(0) \propto \left<I^n\right>$, higher-order nonlinearities will still exacerbate these discrepancies (Fig.~\ref{fig:etanfactorial_adv}(a) and (b)) due to the peak intensity "compression" (Cos) or "selection" (Sin). This difference is also increasingly larger with larger $\xi$ (Fig.~\ref{fig:etanfactorial_adv}(b)), since as explained earlier, large $\xi$ associated with a low MZ non-linearity, select (Sin) or suppress (Cos) the highest peaks $\xi$-times larger than $\langle I_\text{ASE}\rangle$ thus directly impacting in the nonlinear process efficiency.

For $\xi\gtrsim 5$ and sufficiently large $n$, typically $n\ge 8$, the enhancement $\bar{\eta}_n$ exceeds 10${}^3$. As we will discuss later, this is particularly interesting for multiphoton absorption processes in dielectrics such as silica (bandgap energy $E_g=9$~eV) interacting with Yb${}^{3+}$ lasers (photon energy $h\nu=1.2$~eV) which typically results in $n\approx 8$. Describing such interactions solely as an instantaneous nonlinearity is of course a crude approximation, but it highlights a trend that is confirmed with advanced simulations. Moreover it shows that, based on realistic estimates, nonlinearly filtered rare-earth- and more particularly Yb-based ASE sources shall be considered as viable approach for controlling and enhancing light-matter interaction.

Continuously emitting ASE at the required kW-level to obtain strong multiphoton processes ($n>8$) is not essential to produce a source based on the principle described in Section \ref{outline:source description}. Such a power is only required during a short burst of a few ps or ns, and can be easily obtained with an adequate master oscillator -- power amplifier (MOPA) seeded with a pulse-picked ASE source \cite{valero2020}. That is why, from here on all simulations and statistical analysis are performed over a population of $N_p = 10000$ uncorrelated laser pulses of $T_\text{burst}=25$~ps ($1000\cdot\tau_c$). All $N_p$ pulses are statistically independent, as it is the case for a source where the period between two consecutive pulses is larger than $\tau_c$ by many orders.

Pulse-picking a random source has no influence on the statistics of the NLF-ASE such as the PDF, but may greatly influence the energy of an individual sample or the details of the nonlinear dynamics triggered by a particular sample. In order to account for this sample-to-sample fluctuation (see Fig.~\ref{fig:g2gn} in Appendix~\ref{outline:g2gn}), we will keep track of the simulation results for each individual pulse. In addition to ensemble averaging, histograms will thus render not only the average enhancement but also the r.m.s. fluctuations of the physical quantities computed (energy deposition, electron density) as well as the probability of observing the most extreme events in the simulations described in the next sections.

\section{Modeling of the laser energy deposition into a dielectric material}
\label{outline:MRE model}
Considering only instantaneous and unsaturated nonlinearities is the ideal case providing the highest enhancement factor. For more realistic physical systems, there are more complicated cascades, saturated and/or delayed interaction processes which may decrease the enhancement factor. In order to demonstrate the efficiency of the NLF-ASE source for nonlinear excitation within the framework of a more realistic system, the laser energy deposition into a dielectric material is considered.

The phenomenology of the interaction of a short laser pulse focused at the surface of a dielectric material, leading to energy deposition and subsequent material modifications, is as follows \cite{Rethfeld}. First, the valence electrons are promoted to the conduction band through multiphoton absorption or tunneling. The excited electrons may further absorb photons through collisions with phonons, ions, or electrons. In the same time, the conduction electrons may recombine to the valence band or defective states. When the energy of free electrons is on the order of the bandgap, impact ionization may take place: the collision of a free electron with an electron of the valence band leads to two free electrons at the bottom of the conduction band. For long enough laser pulses with sufficiently high intensity, the latter process replicates, leading to the so-called electron avalanche process.

When the free electron density ($n_e$) becomes on the order of the critical plasma density ($n_c$), the laser absorption is significant, leading to an energy density deposited into the material going from hundreds of J/cm$^3$ to hundreds of kJ/cm$^3$ depending on laser parameters. On a 10~ps timescale, the electron energy is transferred to the lattice through electron-phonon and electron-ion collisions. Both the lattice temperature and pressure then grow up, leading to phase transition and material ejection from the surface (ablation). During this interaction process, the time-dependent laser intensity may be modified through laser propagation effects. When $n_e\simeq n_c$, absorption and reflection also significantly take place, changing the spatial intensity profile in the material bulk along the direction of laser propagation. In turn, the production of free electrons may be decreased beneath the material surface.


\begin{figure}[t!]
\centering
\includegraphics[width=\columnwidth]{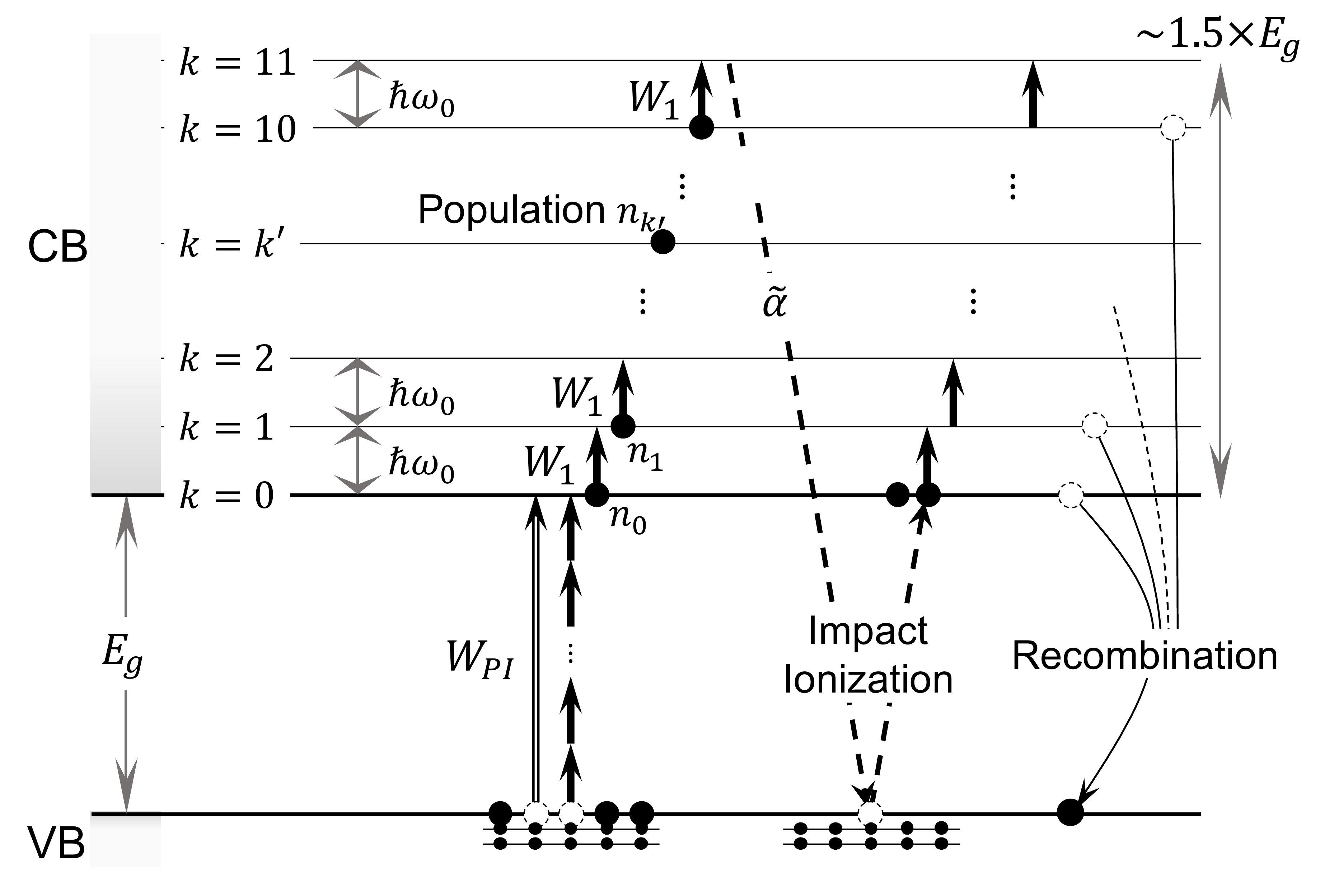}
\caption{Schematic representation of the MRE Model for the electron populations dynamics. The Valence Band (VB) and the Conduction Band (CB) are separated by a bandgap energy $E_g$= 9 eV. The VB is treated as a reservoir of electrons, while the CB is discretized in $k=11$ levels separated by the photon energy $\hbar\omega_0$. Photo-ionization, one-photon absorption, impact ionization, and recombination drive the evolution of the $\{n_k\}$ populations of the $\{k\}$ levels. See main text for the details of notations.
\label{fig:EnergyScheme}
}
\end{figure}


Within the present work which aims at demonstrating the efficiency of the NLF-ASE source in terms of laser-matter coupling, we will choose laser fluences such that the produced free electron density is of the order or less than the critical plasma density. As mentioned below, such a configuration allows us to make assumptions to simplify the modeling. Within this physical condition, a simple and state-of-the-art modeling approach is chosen.

The key physical quantity driving the laser energy absorption is the time-dependent electron density. Since relatively long pulses are considered, a reliable approach to evaluate the electron density is based on multiple rate equations (MRE) driving the electron populations \cite{Rethfeld}. Such an approach has been shown to provide reliable predictions compared with experimental data \cite{Gaudfrin, PRB}. The laser induced electron dynamics described by the present MRE model is illustrated in Fig.~\ref{fig:EnergyScheme}. Within this model, the conduction band is described by a set of allowed states, the energy difference between two adjacent states being the photon energy. All adjacent states are bridged through one-photon absorption, describing the laser heating of conduction electrons. The lowest energy state is filled by both photo- and impact ionization. Finally all states are assumed to decay similarly accounting for the recombination process. The MRE then read
\begin{eqnarray}
\dot{n}_0 & = & W_{PI} + 2 \tilde{\alpha} n_k - W_1 n_0 - n_0 / \tau_r \nonumber \\
\dot{n}_1 & = & W_1 n_0 - W_1 n_1 - n_1 / \tau_r \nonumber \\
& ... & \label{eq:MRE} \\
\dot{n}_{k-1} & = & W_1 n_{k-2} - W_1 n_{k-1} - n_{k-1} / \tau_r \nonumber \\
\dot{n}_{k} & = & W_1 n_{k-1} - \tilde{\alpha} n_{k} - n_k / \tau_r \nonumber,
\end{eqnarray}
where $\dot{n}_i$ is the temporal derivative of the electron density in the $i$ state. $W_{PI}$ is the photo-ionization rate evaluated with the Keldysh expression \cite{Keldysh_2} with a bandgap energy of $E_g = 9$~eV \textcolor{black}{which} is representative of fused silica. $\tilde{\alpha} = 1$~fs is the characteristic time for impact ionization which take place when electrons reach the highest state with $k = 11$. \textcolor{black}{In order to satisfy the energy conservation within the impact ionization process, a conduction electron must have an energy of at least $1.5\cdot E_g$ \cite{Rethfeld}. Since the energy gap between two adjacent levels is the photon energy, the choice of 11 levels ensures that the energy of the highest level is $1.5\cdot E_g$, thus allowing the impact ionization process.} $W_1$ is the one-photon absorption rate in the conduction band set to $3.5\times 10^{-7} E^2$, where $E$ is the laser electric field \cite{Rethfeld}. $\tau_r$ is the recombination time set to $500$~fs or $1$~ps for this study. 

The total electron density is obtained as $n_e(t) = \sum_i n_i(t)$. Note that since cases such that $n_e \lesssim n_c$ will be considered, with $n_c \simeq 10^{21}$~cm$^{-3}$ for a wavelength of $1$~µm, the ionization rates, that depend on the density of valence electrons which is of the order of $10^{22}$~cm$^{-3}$, will not change significantly in the course of interaction. Also note that the used photon energy in the transition rates corresponds to the central wavelength of the laser source, i.e. $1$~µm. Any spectral broadening (a few nm) due to the pulse temporal shape is not included since its influence is negligible for the present processes. 

The absorbed \textcolor{black}{laser} energy \textcolor{black}{results in a deposited energy} density $U$, obtained with
\begin{equation}
\label{eq:abs_las_energy}
\frac{dU}{dt} = k_0 \Im(\hat{n}) I(t),
\end{equation}
which accounts for a laser absorption over a length of the order of the skin depth \cite{Gaudfrin}. $k_0$ and $I$ are the laser pulse wave vector and intensity, respectively. $\Im(\hat{n})$ is the imaginary part of the complex refractive index $\hat{n}$. The latter is evaluated with the Drude model which input parameters are the free electron density $n_e(t)$ evaluated with the MRE model, and the collision frequency which is set to $10$~fs. Within this evaluation of the absorbed energy density, any laser reflection is neglected since $n_e \lesssim n_c$ will stand. This energy absorbed by the electron sub-system is assumed to be eventually transferred to the lattice, which induces the material modifications at the end of the laser pulse.

Since relatively long pulses are considered, of the order of tens of ps, some material changes may occur during the interaction due to the lattice temperature increase, including an evolution of the band structure which may affect the electronic transition rates (ionization rates, collision frequency, recombination time, etc). The objective of the present work is to compare the energy deposition into a dielectric material induced by different laser sources so that such transient effects are not expected to play a significant role on the forthcoming comparative conclusions.

\begin{figure}[t!]
\centering
\includegraphics[width=1.\linewidth]{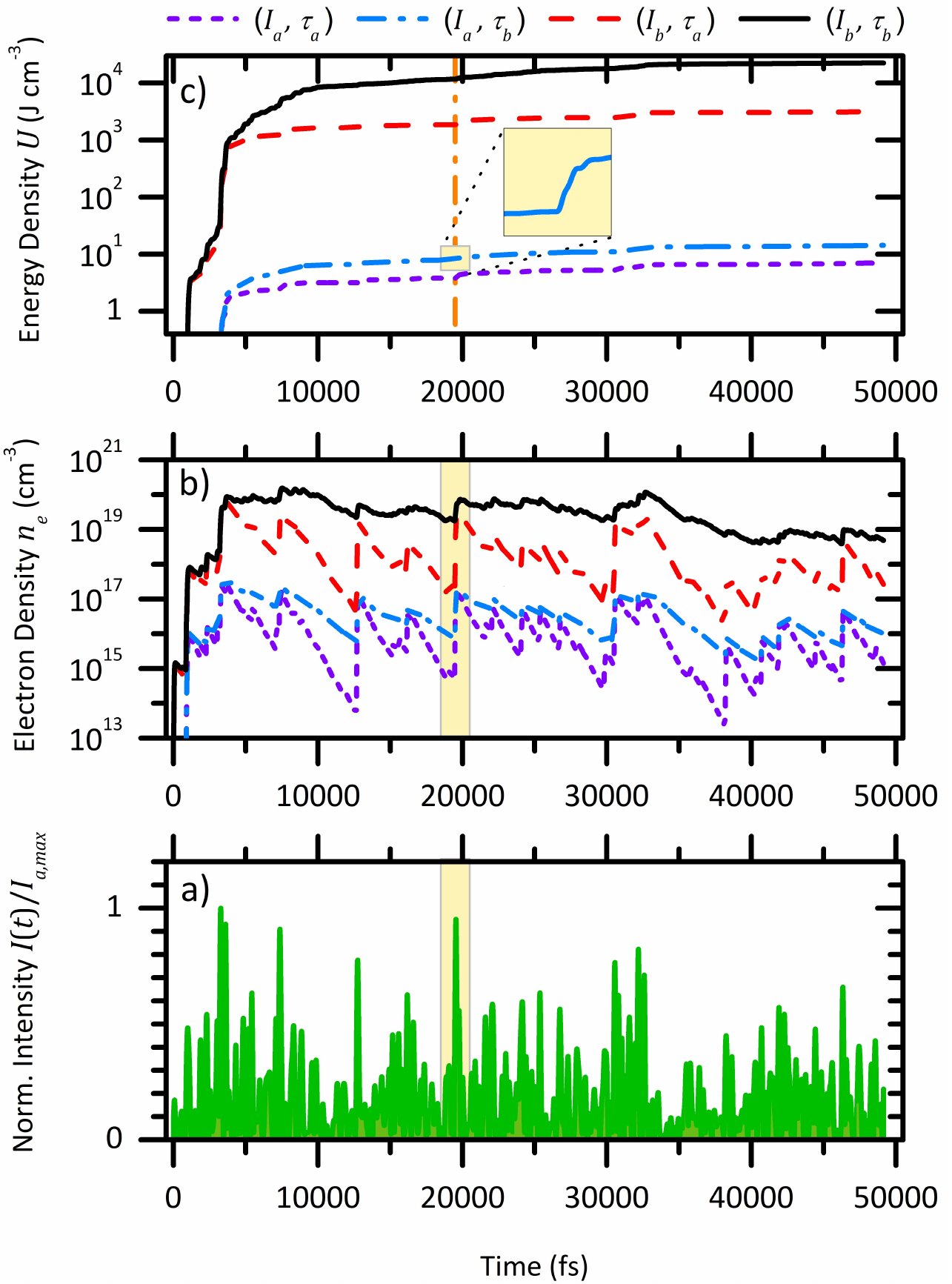}
\caption{(a) Normalized intensity profile of a representative temporal window of a standard ASE laser source. Temporal evolution of (b) the free electron density and (c) the absorbed energy density induced by the present laser intensity profile. Two values of the average intensity and two values of the electron recombination time are used: $I_a = 3$~TW/cm$^2$, $I_b = 7$~TW/cm$^2$, $\tau_a = 500$~fs, $\tau_b = 1$~ps. The legend of the four resulting curves is provided on top of the present figure. The salmon area depicts details of the dynamics around $20$~ps as an example.}
\label{fig:ase}
\end{figure}

All previous material parameters have been set to reasonable values, slight deviations from these values are not expected to change the conclusions of this work. It is worth mentioning that within such conditions, the value of the recombination time in particular is unknown. It is $150$~fs for silica excited by fs laser pulses where the band structure is not expected to change significantly, and it could be longer for other materials or when the material transforms into a plasma \cite{Kar, Pineau}. We thus have chosen a reasonable characteristic recombination time in the ps range. In the forthcoming study, its value is varied to highlight a possible interaction with a memory effect.

\begin{figure}[t!]
\centering
\includegraphics[width=1.\linewidth]{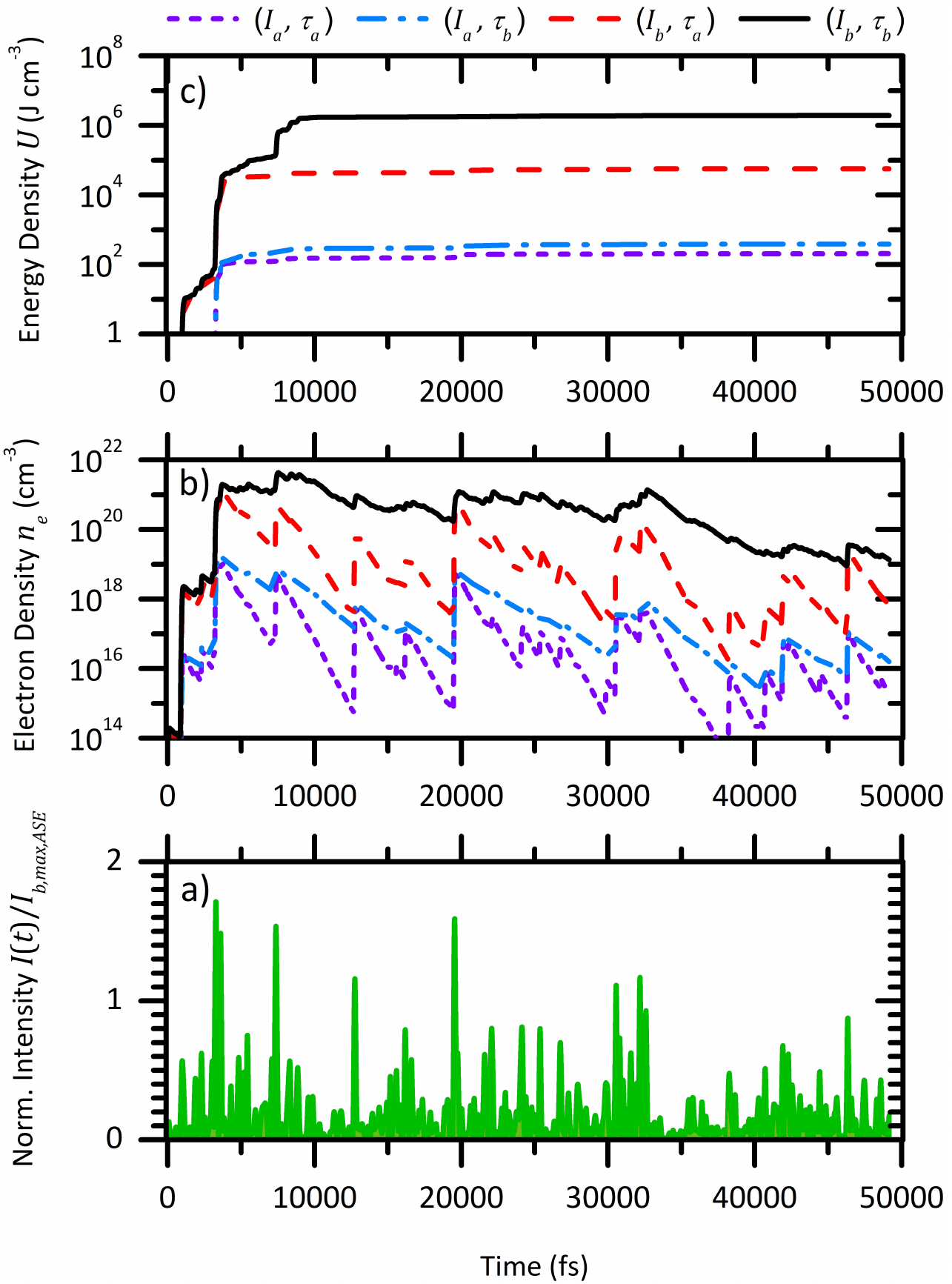}
\caption{(a) Normalized intensity profile of a representative temporal window of the Sin output. This profile is generated from the same ASE profile of Fig. \ref{fig:ase}. It thus shares similar features but is modulated through the NLMZ. Temporal evolution of (b) the free electron density and (c) the absorbed energy density induced by the present intensity profile. Two values of the average intensity and two values of the electron recombination time are used: $I_a = 3$~TW/cm$^2$, $I_b = 7$~TW/cm$^2$, $\tau_a = 500$~fs, $\tau_b = 1$~ps. The legend of the four resulting curves is provided on top of the present figure.}
\label{fig:sb}
\end{figure}

\section{ASE vs. NLF-ASE energy deposition - results and discussion}
\label{outline:silica absorption}

Because of their random nature, both ASE and NLF-ASE exhibit complex temporal structures. The highly nonlinear interaction of such stochastic light sources with a dielectric material within the MRE model further increases the complexity of the process, due to the possible cumulative effects of numerous low-intensity peaks (acting as a quasi-continuous background) and strong intensity spikes with random time-ordering.

In order to gain some insight into the dynamics of energy deposition obtained from full simulation, and to disentangle the influence of time-ordering or peak-to-background contrast from the random aspects, instead of ASE or NLF-ASE, "toy" pulses can be used (see Appendix \ref{sec:toy}).

These simple models show the main variations of the energy density are due to the intensity spikes that are greatly favoured by nonlinear filtering. Quasi-continuous background following a spike contributes to the energy deposition as well.

The temporal evolution of both free electron density and energy deposition induced by ASE and NLF-ASE laser sources is first investigated. For illustration purposes, a representative $T_\text{burst}=50$~ps temporal window of the intensity profile of these sources is chosen. Statistical variations of electron and energy density are then addressed by considering a large number of bursts with the same temporal windows but random intensity profiles.

\textcolor{black}{The data presented in Figs.~\ref{fig:ase} and \ref{fig:sb} are obtained from the model presented in Section \ref{outline:MRE model} and illustrated by Fig.~\ref{fig:EnergyScheme}. In practice, Eqs.~(\ref{eq:MRE}) and (\ref{eq:abs_las_energy}) are numerically solved to obtain the free electron density, and electron energy density, respectively. A standard finite difference method is used to solve the differential equations. Hereafter, The "energy density" then corresponds to the electrons' energy.}

Fig. \ref{fig:ase} depicts the temporal evolution of both free electron density and energy density as induced by a time-window of a standard ASE laser source. The normalized intensity profile is shown in Fig. \ref{fig:ase}(a). To study the influence of the average intensity or fluence, two values are chosen ($I_a = 3 $~TW/cm$^2$ and $I_b = 7 $~TW/cm$^2$) which correspond to an experimentally achievable laser power of the order of $10$~W (with a 10 µm focal spot size and a repetition rate of $100$~kHz for instance). Regarding the electron recombination time, an influence on the whole interaction is expected when it varies in the range of intensity peak-to-peak spacing out, i.e. of the order of $1$~ps. $\tau_a = 500$~fs and $\tau_b = 1$~ps are thus chosen.

As shown by Fig. \ref{fig:ase}(b), the electron dynamics mainly consists of sharp increases and slower decreases, due to the highest intensity spikes (single events) and regions of moderate intensities, respectively. This behavior is illustrated by the filled salmon area around $20$~ps. At first the electron density is smoothly decreasing due to the electron recombination (the intensity is too low for the production photoionization rate to compensate the electron recombination loss). Then the intense spike at $\sim 20$~ps leads to a sharp increase in the electron density owing to the nonlinear feature of the interaction as presented in Appendix \ref{sec:toy}. Considering the whole time window, as expected the higher the average intensity, the larger the free electron density. Regarding the influence of the recombination time, after an intensity spike, the longer this time, the slower the decrease in electron density, leading to the final largest electron density with the longest recombination time. However its influence depends on the average laser intensity.

For the lowest intensity, regardless of the value of the recombination time, the shape of the electron density is similar, i.e. the decrease after a spike is always significant. For the highest average intensity, in the case of the longest recombination time, the ionization rate is large enough to counterbalance the recombination loss in the regions between spikes, leading to a significantly lower decrease and a change in the temporal shape of the electron density which becomes almost constant after the main production event around 3300~fs.

Since the energy deposition is proportional to the cumulative sum of the free electron density, it always increases monotonically with respect to time regardless of the values of parameters (no loss mechanism is included on this short timescale). The variations of the final energy deposition with respect to both intensity and recombination time mimic the behavior of the electron density. The higher the average laser intensity and the longer the recombination time, the larger the laser energy deposition into the material. Note that as shown in Appendix \ref{sec:toy}, the main variations of the energy density are due to the intensity spikes, as illustrated by the inset in Fig.~\ref{fig:ase}(c).

Fig.~\ref{fig:sb} depicts the temporal evolution of both free electron density and energy density as induced by a time-window of the NLF-ASE laser source. All parameters are the same as for results presented in Fig.~\ref{fig:ase} for the ASE source. All previously observed trends are overall retrieved in that case. The main difference consists of sharper variations due to the intensity spikes which are more intense for the NLF-ASE source compared with the ASE source. As a result, both the produced free electron density and energy deposition are larger with the NLF-ASE source (for the same total fluence).

The previous results highlight the main behavior of the interaction dynamics with the NLF-ASE and ASE laser sources. They have been obtained for a single temporal window. Since these sources are likely to be pulse-picked each pulse exhibit shot-to-shot fluctuations. Although  the individual pulse statistics and overall statistics are identical \footnote{We numerically checked that time- and ensemble averaging lead to the same results so that the $N_p$ PDFs of $N_p$ independent pulses of duration $T_\text{burst}$ are identical to (i) the mean PDF averaged over the $N_p$ individual PDFs, and also to (ii) the PDF of a $N_p\cdot T_\text{burst}$ longer sample.}, each sample has a different spike intensity distribution in time, and an individual energy that slightly fluctuates too. A statistical evaluation of the laser energy deposition following the interaction process should thus be performed. In order to render the distribution of the  average density of free electrons produced and associated final energy density, a large number of temporal windows set to $N_p=10000$ is considered in the present work. 
 
\textcolor{black}{As discussed earlier in Section \ref{outline:AnalyticalPDF}, the exact simulation of the laser propagation in the NLMZ fibers can be replaced by using Eqs.~(\ref{eq:Sin intensity}) and (\ref{eq:Cos intensity}) instead.
Indeed, for the value $\xi=5$ chosen here and leading to a large enhancement, the nonlinear phase shift induced in the NLMZ does not cause any tangible spectral broadening. Consequently the dispersion does not spread the sub-pulses of $I_\text{ASE}(t)$ in time, and both the peak structure and the PDF are identical to those described in Eqs.~(\ref{eq:Sin intensity})-(\ref{eq:p_s alpha}), (\ref{eq:composition PDFs deriv}) and (\ref{eq:composition PDFs deriv 2}). The evaluation of the intensity profiles $I_\text{ASE}(t)$, $I_s(t)$, and $I_c(t)$ for each random sample is thus numerically tractable. For a smaller number of temporal windows, we have checked that the obtained intensity PDFs are similar to those obtained by a full calculation using an NLSE solver \cite{fiberdesk}. The corresponding PDF (intensity spikes distribution) is provided in Fig. \ref{fig:statSources}(c).}

\begin{figure}[t!]
\centering
\includegraphics[width=1.\linewidth]{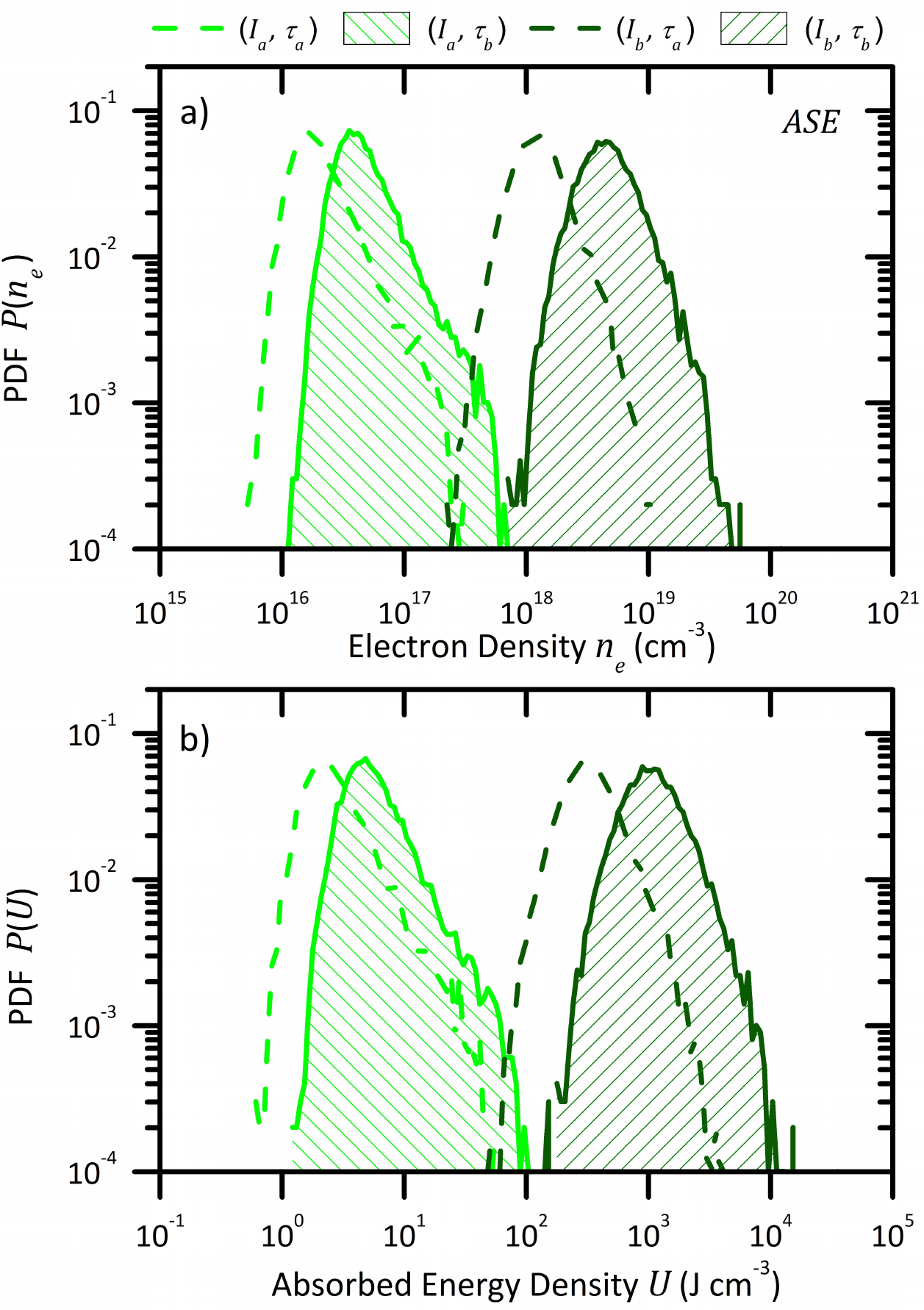}
\caption{\textcolor{black}{Probability} Distribution \textcolor{black}{Function (PDF)} of \textcolor{black}{(a)} free electron density and \textcolor{black}{(b)} absorbed energy density induced by an ASE source. 10,000 temporal windows have been used to construct the distributions. The simulation parameters (mean intensity, recombination time) are the same as previous cases.}
\label{fig:statASE}
\end{figure}

\begin{figure}[ht!]
\centering
\includegraphics[width=1.\linewidth]{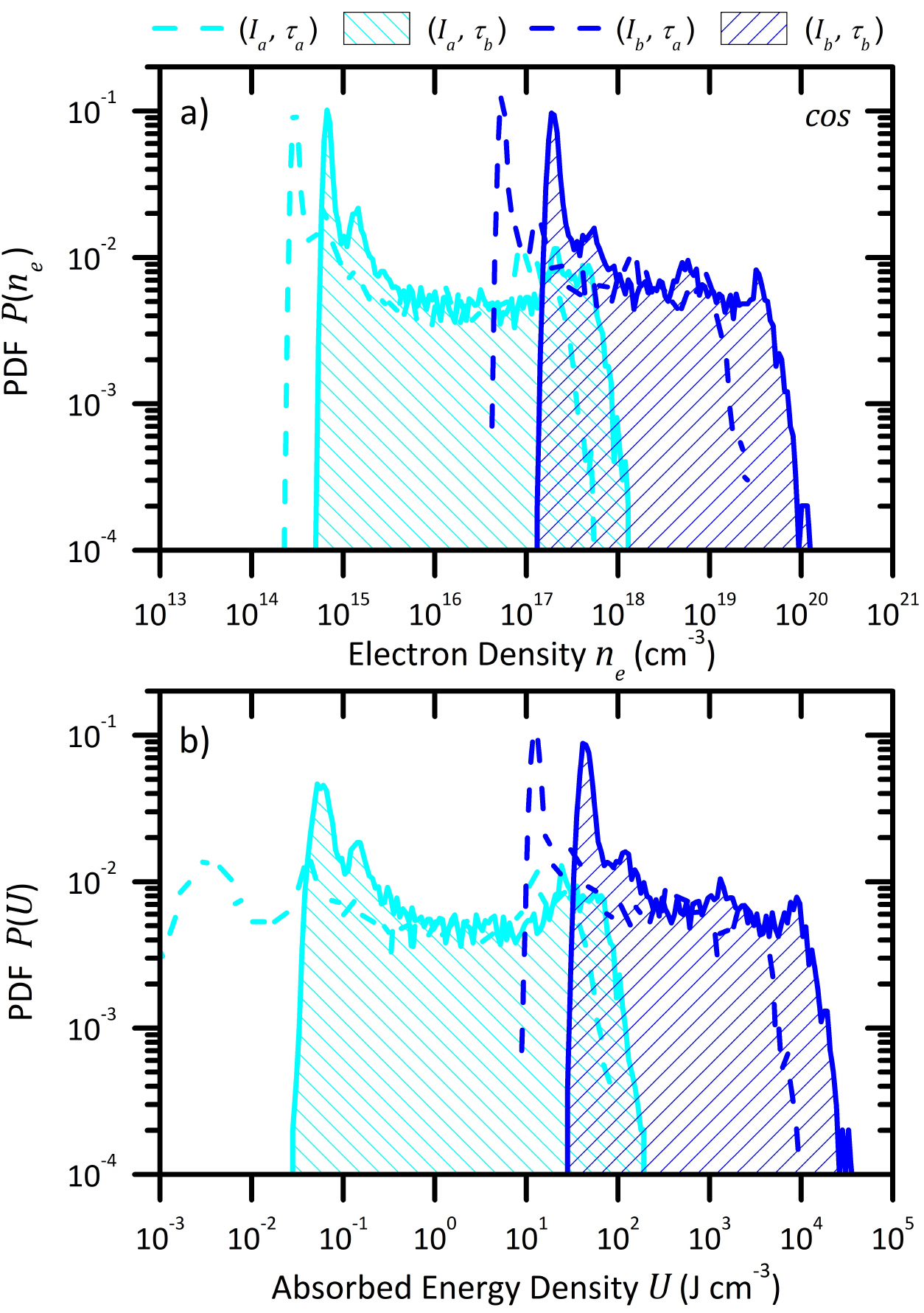}
\caption{\textcolor{black}{Probability} Distribution \textcolor{black}{Function (PDF)} of \textcolor{black}{(a)} free electron density and \textcolor{black}{(b)} absorbed energy density induced by an NLF-ASE(cos) source. The simulation parameters (10,000 temporal windows, mean intensity and recombination time) are the same as previous cases.}
\label{fig:statcos}
\end{figure}

\begin{figure}[ht!]
\centering
\includegraphics[width=1.\linewidth]{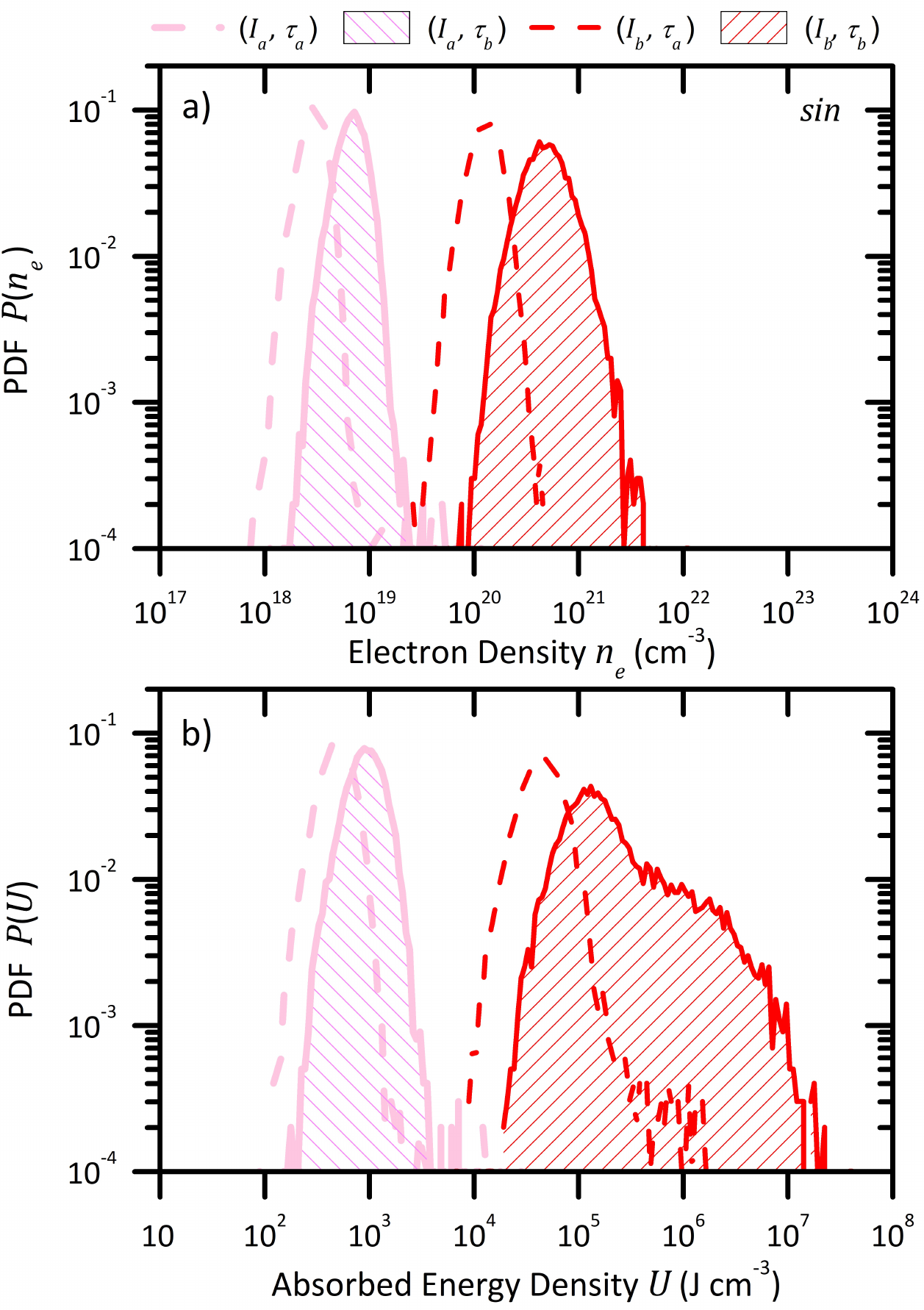}
\caption{\textcolor{black}{Probability} Distribution \textcolor{black}{Function (PDF)} of \textcolor{black}{(a)} free electron density and \textcolor{black}{(b)} absorbed energy density induced by an NLF-ASE(sin) source. The simulation parameters (10,000 temporal windows, mean intensity and recombination time) are the same as previous cases.}
\label{fig:statsin}
\end{figure}

The distribution in average electron density and energy density induced by the ASE, NLF-ASE Cos and Sin sources are shown in Figs. \ref{fig:statASE}, \ref{fig:statcos}, and \ref{fig:statsin}, respectively. Values of average intensity and recombination time are similar as previously. For the ASE source, overall, the distributions are rather symmetric around a certain value, exhibiting a shape with tails close to a power law. The previously mentioned influence of parameters is retrieved.

A higher average intensity leads to a distribution centered on larger electron and energy densities. There are two orders of magnitude between mean electron and energy densities produced with the presently used two average intensities, which roughly corresponds to $2^8$, accounting for the signature of the $8$-photon absorption process to ionize the material. Note that the distributions exhibit a bump on their right part. It should be related to the extreme events of highest sparse intensity spikes which further produce a large amount of free electrons through nonlinear absorption.

The distribution shape induced by the NLF-ASE Cos source is dramatically different as shown by Fig. \ref{fig:statcos}. It consists roughly of a main peak with a plateau-like for larger values of electron or energy density. It is due to the distribution of intensity spikes of the Cos source of which the shape is significantly different from the ASE one. Instead of exhibiting a long tail, it is rather constant with a sharp drop at $2\left<I\right>$. It follows that window-to-window fluctuations are less expected, thus leading to the observed narrow peak in the distribution of electron and energy densities. The plateau-like behavior should be due to the increase in the intensity spike distribution before dropping. 

Regarding the most probable values of electron and energy densities, for a given set of parameters, they are smaller than those induced by the ASE source. Indeed the Cos source exhibits a more homogeneous profile in time than the ASE source as depicted in Fig.~\ref{fig:statSources}(b), with significantly less extreme intensity spike events which are responsible of large free electron production.

The distributions in electron and energy density induced by the NLF-ASE Sin source are provided in Fig.~\ref{fig:statsin}. Their main shape is similar to the one induced by the ASE source. It is due to the fact that the intensity spike distribution of both sources exhibits similar features: it mainly consists of a monotonic decrease over a large intensity range. However the mean electron and energy densities induced by the NLF-ASE Sin source are larger than those induced by the ASE source. As mentioned above, the tail of the NLF-ASE Sin source distribution expands to higher amplitudes of intensity spikes than the ASE one, explaining the observed trends again owing to the nonlinear absorption.

\begin{figure}[t!]
\centering
\includegraphics[width=1.\linewidth]{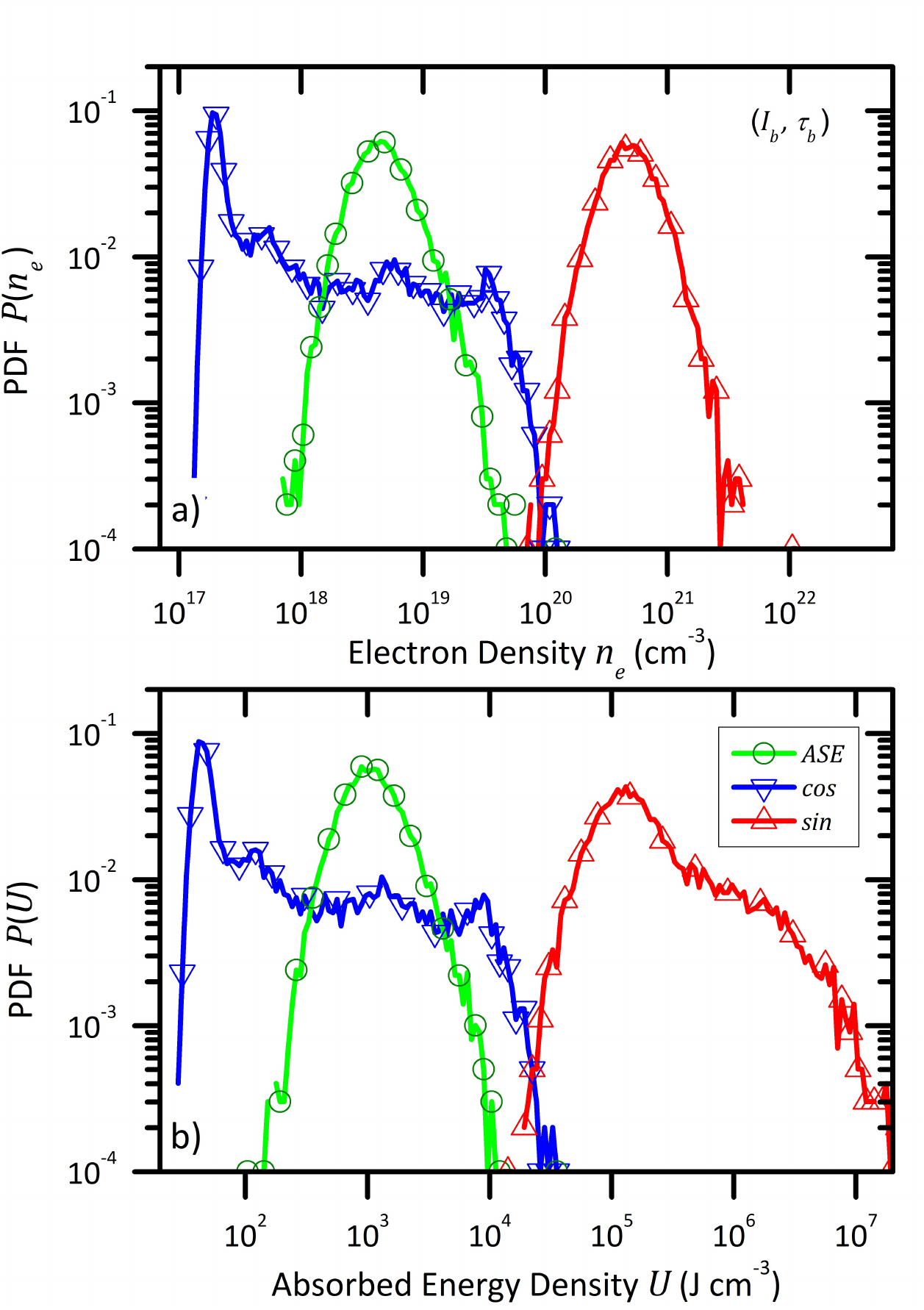}
\caption{Direct comparison of the \textcolor{black}{Probability} Distribution \textcolor{black}{Function (PDF)} of \textcolor{black}{(a)} free electron density and \textcolor{black}{(b) absorbed} energy density for the three laser sources where the optimal parameters are used (highest intensity \textcolor{black}{$I_b$} and longest recombination time \textcolor{black}{$\tau_b$}).}
\label{fig:statOptimal}
\end{figure}

Overall, due to the nonlinear feature of the present interaction, the Sin NLF-ASE source leads statistically to both the highest free electron density and energy deposition into the material by a factor of $100$ in average (a direct comparison is provided in Fig.~\ref{fig:statOptimal} with parameters leading to the largest energy deposition). Despite the fact that this enhancement factor is smaller than the above mentioned ideal nonlinear response of a physical system, it remains relatively large for the presently studied realistic physical system, demonstrating the interest of the proposed NLF-ASE source for nonlinear interaction.

Although the NLF-ASE Cos case is less efficient in terms of laser energy deposition into a dielectric material, it offers the possibility of a larger range of energy deposition and subsequent large diversity of material modifications and applications. Since the transition from Cos to Sin distributions can be simply tuned with a single parameter, such a NLF-ASE laser source is an interesting tool which can adapt to the desired application. For instance for laser surface structuring where the focal spot is moved along the material surface to process some area, various surface states and properties could be achieved by tuning the distribution of the intensity spikes.  
\section{Final discussion}
\label{outline:discussion}

NLF-ASE sources offer an interesting alternative to the already existing superbunched sources. Our simulations suggest that they are accessible using current fiber components and Yb${}^{3+}$ laser setups. Compared to other types of stochastic sources, they may even achieve a much higher optical throughput, characteristics that may be of high potential for the studies of nonresonant and nonlinear processes excited by stochastic sources.

The analytic derivations presented here show that the free parameters $\varphi$ and $\xi$ allow to adjust finely the intensity PDF of the source. The PDF may be anticipated owing to the derived equation Eq.~(\ref{eq:p_s alpha}). Among other criteria, different sets of parameters will give different background/peak energy ratios, which can have in turn a significant influence on the phenomena investigated. The proposed NLF-ASE source may thus be an efficient tool to study in details the optical response of physical systems.

As an example, the potential of superbunched NLF-ASE sources is clearly demonstrated for nonlinear dielectric absorption in Section \ref{outline:silica absorption}, where the superbunching of the Sin source is shown to trigger a hundredfold enhancement in terms of deposited energy compared to the ASE source. Of course, the interaction model used here includes assumptions, that's why the simulated absorbed energy densities should be considered relatively to one another rather than as absolute values. This numerical investigation thus demonstrates the interest of NLF-ASE sources in experiments involving high-order nonlinear processes with complex dynamics.

The Cos source may be qualified as "antibunched" and is relatively less studied in the present work. However, it may also have very interesting applications in a near future. Indeed, some high-power laser experiments require a low-coherence-time source with, ideally, nonlinear effects substantially lowered compared to that of e.g. natural ASE sources. That is exactly what the Cos source is capable of. \textcolor{black}{More generally, to each nonlinear order $n$ corresponds a range in $\xi$ and $\varphi$ parameters giving antibunched sources, that is, sources for which $\eta_n$ is appreciably inferior to 1. Taking e.g. both $\xi = 4.2$ and $\varphi = 0.07$~rad,} we achieved numerically $g^{(2)}(0)$ values down to \textcolor{black}{1.4}  -- close to the theoretical limit of 1 for classical light -- instead of 2 for ASE. Such a feature could be of high importance for instance in facilities dedicated to inertial confinement fusion where the use of stochastic sources dawns as a possible alternative to coherent front-ends \cite{cui2019high} in order to tame parametric instabilities in the interaction plasma as well as in the laser amplifiers. 

The systematic exploration of the nonlinear Mach-Zehnder parameters $(\varphi, \xi)$ and their influence on both optical statistics and nonlinear processes proves time-consuming and is beyond of the scope of this work. Let us note however that some points have shown unexpected features, such as an antibunching for some nonlinear orders (e.g. $n < 5$) and a superbunching for others (e.g. $n \geq 5$). Such configurations would be very promising for applications where the propagation in the laser chain is a concern because of low-order instabilities, while aiming at triggering  efficient higher-order phenomena, e.g. multiphoton absorption. This vast horizon of possible outcomes sketches a complex but all the more promising field of study. It is also a strong incentive to undertake the physical implementation of a NLF-ASE source.

\begin{figure}[b!]
\centering
\includegraphics[width=1.\columnwidth]{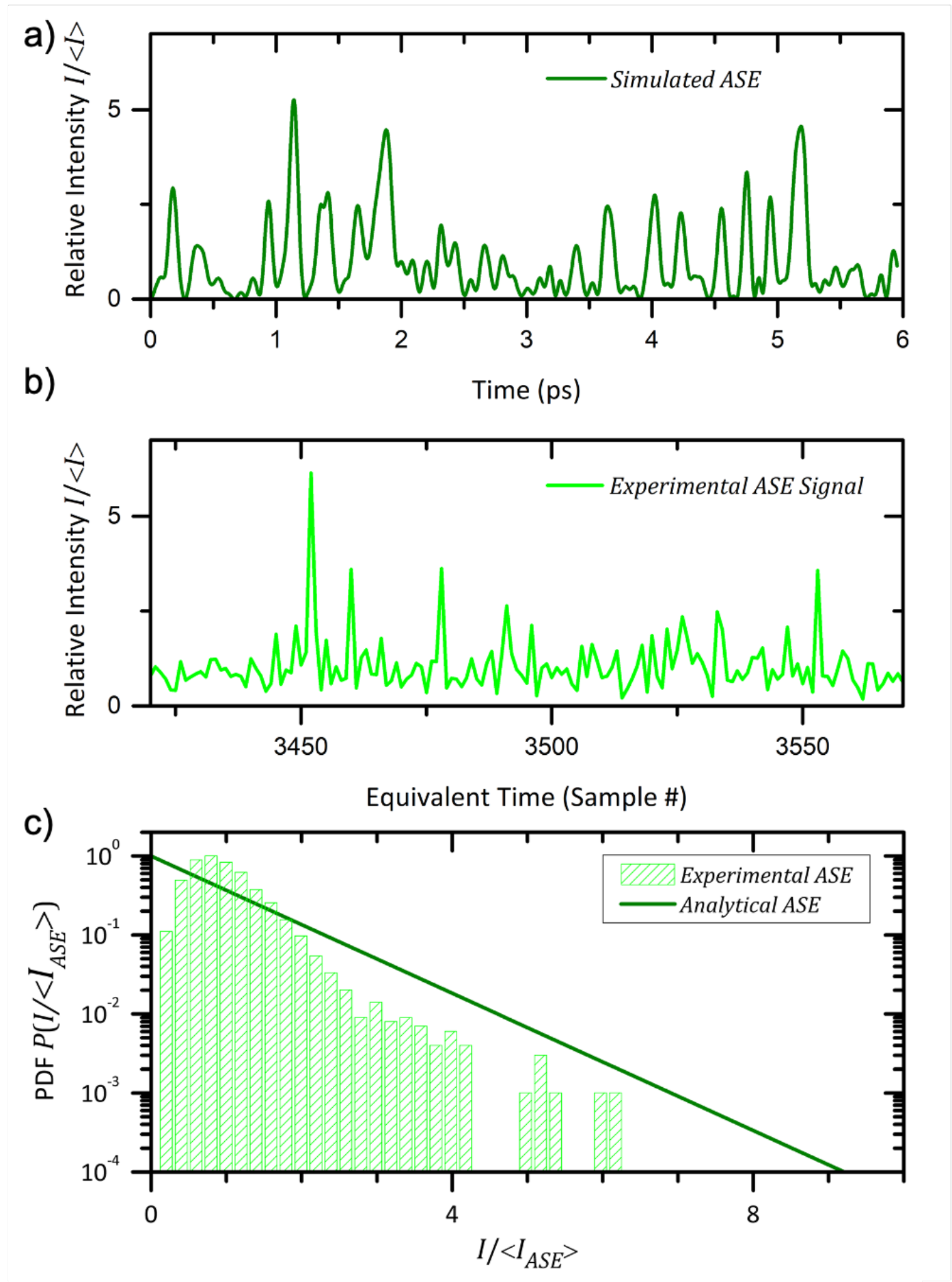}
\caption{Time traces and Probability Density Function of ASE: a) Simulated normalized intensity profile of an ASE source, b) Experimental normalized intensity profile of an ASE source. c) PDF of signal a) and b).}
\label{fig:AppendixASE}
\end{figure}

\appendix

\section{Simulation and basic properties of ASE}
\label{outline:appendix ASE}
The ASE model used in parts \ref{outline:nlMZ simulation} and \ref{outline:silica absorption} is identical to that of \cite{valero2020}. It is modeled as a purely Markov-Gaussian process by generating a random spectrum. 

We generate a vector $S$ representing the spectrum of the signal centered at 1030~nm. The spectrum is then generated by drawing random numbers for the phase (uniform distribution between 0 and 2$\pi$) and amplitude (uniform distribution between 0 and 1) on a spectral window $\Delta f$. All other modes are canceled out. $\Delta f$ may vary between 0 and $\Delta f_\text{max}$ chosen such as to avoid spectral folding. The parameters used for this work (bandwidths about 40~nm) are typically much smaller than $\Delta f_\text{max}$.

The complex (non-dimensional) field $E(t)$ is obtained by carrying out a Discrete Fourier Transform (DFT) on $S$. The time increment is typically of 1~fs. Afterwards, we calculate the time-domain intensity vector $I(t)$ as the square modulus of $E(t)$. 
Fig.~\ref{fig:AppendixASE}(a) shows a short snapshot of the time output of the model, and Fig.~\ref{fig:AppendixASE}(c) illustrates the almost purely Gaussian nature of the simulated ASE as a continuous random variable.

For an Yb${}^{3+}$ source with a relatively narrow-band spectrum \cite{valero2020}, we can observe these fluctuations experimentally with an ultrafast photodiode as shown in Fig.~\ref{fig:AppendixASE}(b). The experimental PDF Fig.~\ref{fig:AppendixASE}(c) is retrieved by a frequency count (normalized histogram) of the trace Fig.~\ref{fig:AppendixASE}(b). Bandwidth, linearity, and noise threshold of the photodiode can rapidly limit the fidelity of the retrieved PDF, but here the agreement is good as shown in Fig.~\ref{fig:AppendixASE}(c). In order to grasp the rarest peaks, long acquisition times are needed, and as shown in Fig.~\ref{fig:AppendixASE}(c), obtaining just enough statistics to observe the onset of $I/\langle I \rangle \simeq 10$ peaks would require 100 to 1000 times more data points, as these relative intensities are conversely less probable than $I/\langle I \rangle \simeq 6$ obtained.   

\section{Analytical derivation of the PDF of NLF-ASE sources -- Comparison with NLSE simulations}
\label{outline:appendix PDF NLF-ASE}

We derive here the probability density function (PDF) of the random photon flux $I_s$ emitted by the Sin output of an NLF-ASE source. Following Fig.~\ref{fig:statSources}(c) and using the notations in the main article, we consider an NLF-ASE source of average intensity $i_0=\left<I_\text{ASE}\right>$ defining the critical parameter $\xi$. The PDF of $I_s$ is hereafter called $p_{s,\xi}(I_s)$. Neglecting optical losses and all nonlinear couplings in the fibered Mach-Zehnder except nonlinear phase accumulation, we write $I_s$ as the intensity transmitted  in $\sin^2$ by the nonlinear Mach-Zehnder:
\begin{equation}
\label{eq:I_s def}
I_s = \sin^2{\left(\frac{\pi}{2}\frac{i}{\xi \cdot i_0}\right)}\cdot i = g_{s,\xi}(i).
\end{equation}
Although not injective, $g_{s,\xi}$ is however monotonous over each $[\mathcal{I}_k,\mathcal{I}_{k+1}]$ interval, where $\mathcal{I}_{2k}$ is the $(k+1)$-th zero of $g_{s,\xi}$ and $\mathcal{I}_{2k+1}$ its $(k+1)$-th maximum. Let us define $\mathcal{I}_{2k} = 2k \xi i_0$. Thus, $g_{s,\xi}$ admits an inverse function locally on each interval $[\mathcal{I}_k,\mathcal{I}_{k+1}]$. Let Pr be the probability of an event to occur, for any value of $I_s$ one can write
\begin{equation}
\label{eq:proba cond}
\text{Pr}\{I_s \in [I_s, I_s+dI_s]\} = \sum_{k=0}^{+\infty} \text{Pr} \{i \in [i_k, i_k+di_k]\} ,
\end{equation}
where $i_k$ is defined by $g_{s,\xi}(i_k) = I_s$ and $i_k \in [\mathcal{I}_k,\mathcal{I}_{k+1}]$. On such intervals where no value $i_k$ may be defined this way, we deem the associated probability term in Eq.~(\ref{eq:proba cond}) equal to 0. Following the theorem of composition of PDFs, we may rewrite Eq.~(\ref{eq:proba cond}) as
\begin{subequations}
    \begin{align}
        p_{s,\xi}(I_s) &=  \mathlarger{\sum}_{k=0}^{+\infty}\frac{p_\text{ASE}(i=i_k)}{\left|\displaystyle \frac{dg_{s,\xi}}{di}(i = i_k)\right|} \label{eq:composition PDFs} \\
        &=\frac{1}{i_0} \mathlarger{\sum}_{k=0}^{+\infty}\frac{e ^{-\frac{i_k}{i_0}}}{\displaystyle\left|\sin^2{\theta_k}+2\theta_k\sin{\theta_k}\cos{\theta_k}\right|} 
        \label{eq:composition PDFs deriv} ,
    \end{align}
\end{subequations}
where $\theta_k = \pi i_k /2 \xi  i_0$.

We can see in Eq.~(\ref{eq:composition PDFs}) that the poles $P_k$ of $p_{s,\xi}$ are the values of $I_s$ for which $g'_{s,\xi}(i_k) = 0$. This corresponds to the definition of $\mathcal{I}_k$. Hence, they may be derived from Eq.~(\ref{eq:I_s def}) as $g_{s,\xi}(\mathcal{I}_k)$:
\begin{align}
    \label{eq: p_s poles zero}
    &P_0 = g_{s,\xi}(\mathcal{I}_{2k}) =0 ,\\
    \label{eq: p_s poles odd}
    &P_{k} = g_{s,\xi}(\mathcal{I}_{2k+1}) =\sin^2{\frac{\pi \mathcal{I}_{2k+1}}{2\xi i_0}}\cdot \mathcal{I}_{2k+1} .
\end{align}
Deriving $g'_{s,\xi}(\mathcal{I}_{2k+1}) = 0 $  we also find the relation
\begin{equation}
\label{eq: equation I_2kp1}
\tan{\frac{\pi\mathcal{I}_{2k+1}}{2\xi i_0}} =-\frac{\pi\mathcal{I}_{2k+1}}{\xi i_0}.
\end{equation}
Eq.~(\ref{eq: p_s poles zero}) shows that $I_s = 0$ is a pole of $p_{s,\xi}$. Also, a simple numerical approximation of the solutions of $\tan{x} = -2x$ injected in Eq.~(\ref{eq: p_s poles zero}) proves that the approximation $P_{k>0}\approx \mathcal{I}_{2k+1} \approx (2k+1)\xi i_0$ induces a relative error smaller than 5\% for $P_1$ and 1\% for all $P_{l\geq 2}$.

Note that, being derived as a PDF of a random variable, $p_{s,\xi}$ remains positive and summable on all intervals comprising a pole. Roughly, one can evaluate by examining Eq.~(\ref{eq:composition PDFs deriv}) that if $\delta I_s \ll \xi i_0 /\pi$,
\begin{equation}
\label{eq:integrale pole}
\text{Pr}\{I_s \in [P_k, P_k + \delta I_s]\} \propto e^{-(2k+1)\xi} \cdot \delta I_s .
\end{equation}

\begin{figure}[t!]
\includegraphics[width=\columnwidth]{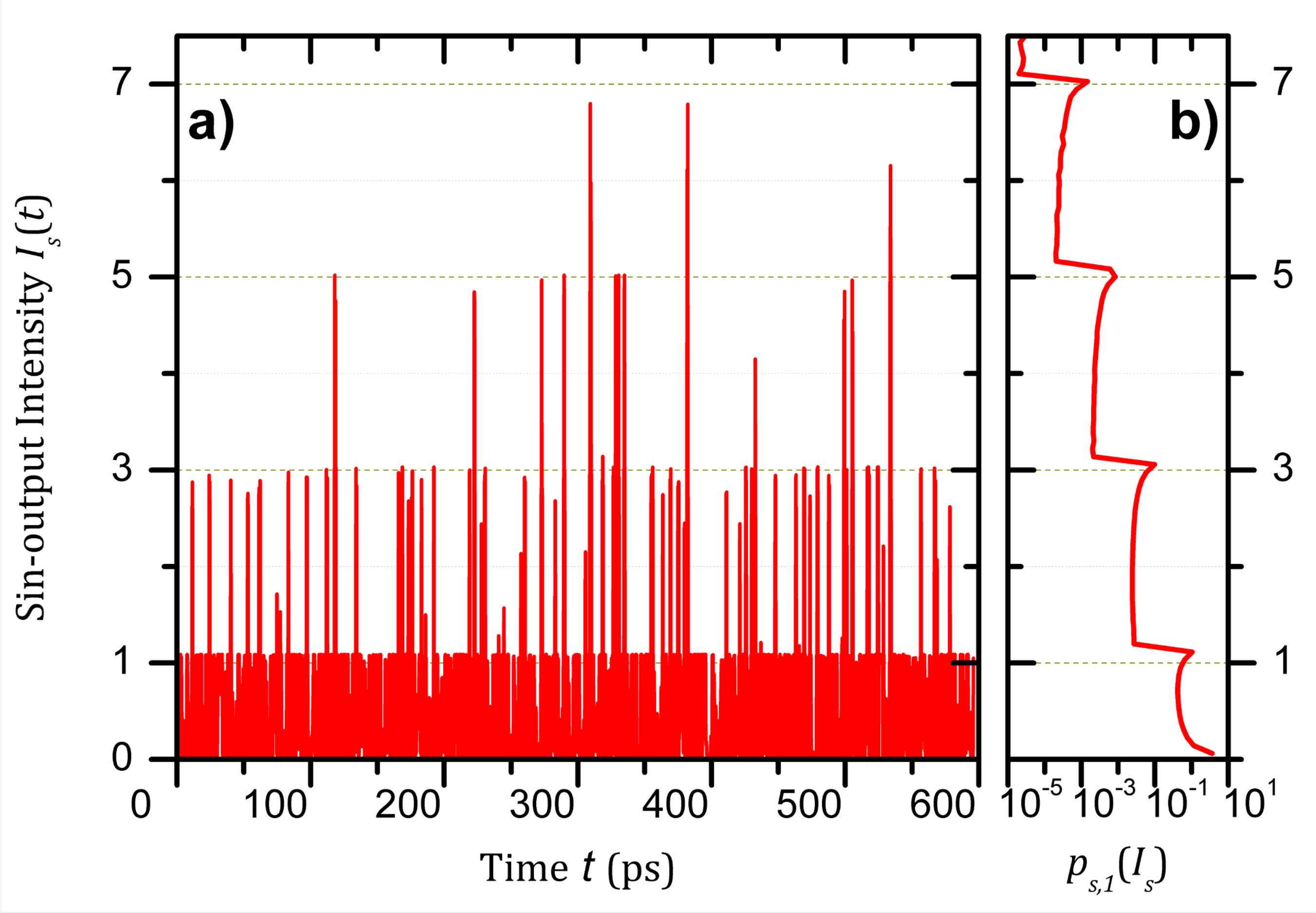}
\caption{
\label{fig:sim_temp_adv}
a) Time simulation (sample) of the Sin output emission of an NLF-ASE source with $i_0 = 1$ and $\xi = 1$. b) PDF (frequency count) of the whole time simulation.}
\end{figure}

Eq.~(\ref{eq:integrale pole}) gives an estimation of the relative probability to obtain a value of $I_s$ around different poles. Finally, let note that when $I_s$ crosses a pole $P_k$, the $(2k)$-th and the $(2k+1)$-th term of Eq.~(\ref{eq:composition PDFs deriv}) vanish, which explains the "quantized" aspect of the temporal emission in Fig.\ref{fig:sim_temp_adv}(a). For $i_0 = 1$ and $\xi = 1$, this quantification is clearly seen as one can clearly distinguish accumulation values around 1, 3, 5 and 7 on the PDF Fig.~\ref{fig:sim_temp_adv}(b), consistent with $P_k \approx (2k+1)\xi i_0$.

The derivation of the PDF of the Cos port, $p_{c,\xi}$, follows closely that of $p_{s, \xi}$. Using $g_{c,\xi}(i)=i\cdot\cos^2{\left(\pi i/2\xi i_0\right)}$ it reads
\begin{subequations}
    \begin{align}
        p_{c,\xi}(I_s) &=  \mathlarger{\sum}_{k=0}^{+\infty}\frac{p_\text{ASE}(i=i_k)}{\left|\displaystyle \frac{dg_{c,\xi}}{di}(i = i_k)\right|} \label{eq:composition PDFs 2} \\
        &=\frac{1}{i_0} \mathlarger{\sum}_{k=0}^{+\infty}\frac{e ^{-\frac{i_k}{i_0}}}{\displaystyle\left|\cos^2{\theta_k}-2\theta_k\sin{\theta_k}\cos{\theta_k}\right|} 
        \label{eq:composition PDFs deriv 2} .
    \end{align}
\end{subequations}

The analysis carried out for $p_{s,\xi}$ remains the same for $p_{c,\xi}$, with poles $P_k(\xi) = g_{c,\xi}(\mathcal{I'}_k)$ now given by the relation
\begin{equation}
\label{eq: equation I_2kp1 cos}
{\cot\frac{\pi\mathcal{I'}_{2k+1}}{2\xi i_0}} =\frac{\pi \mathcal{I'}_{2k+1}}{\xi i_0}.
\end{equation}

The comparison of Eqs.~(\ref{eq:composition PDFs deriv}) and (\ref{eq:composition PDFs deriv 2}) shows that the photon statistics of the two ports of the nonlinear Mach-Zehnder (NLMZ) interferometer are essentially identical in nature. However, as stated by Eqs.~(\ref{eq: equation I_2kp1}) and (\ref{eq: equation I_2kp1 cos}), their poles are completely distinct, leading to very different PDFs. This explains why their merits in terms of nonlinear interaction may differ by many orders of magnitude as simulated in Section \ref{outline:silica absorption}.

The actual propagation in the nonlinear elements of the NLMZ involves chromatic dispersion that has been neglected in the above derivation of the PDF. For sake of comparison, the intensity profiles of both Sin and Cos output have been computed as follows.
An initial ASE signal is generated according to the method described in Appendix~\ref{outline:appendix ASE}. A total of $2^{20}$ points are generated on a time interval of 3.2~ns corresponding to $\sim 3$~fs resolution. The total energy is 5~µJ thus leading to an average power of $P_\text{ASE}=$1.5 kW. Due to the 50:50 splitting ratio, each arm of the NLMZ receives 2.5~µJ (750~W of average power). Assuming $\alpha_2=0.2$ (a 5:1 ratio between the two arms) in order to introduce a differential nonlinearity, the energy is 2.5~µJ ($P_1=P_\text{ASE}/2=750$~W) in arm (1), and  0.5~µJ in arm (2) ($P_2=\alpha_2 P_\text{ASE}/2=150$~W). For these energies and the parameters listed below, the critical NLMZ parameter (Eq.\ref{eq:crit parameter def}) is $\xi\simeq5$.

\begin{figure}[t!]
\includegraphics[width=\columnwidth]{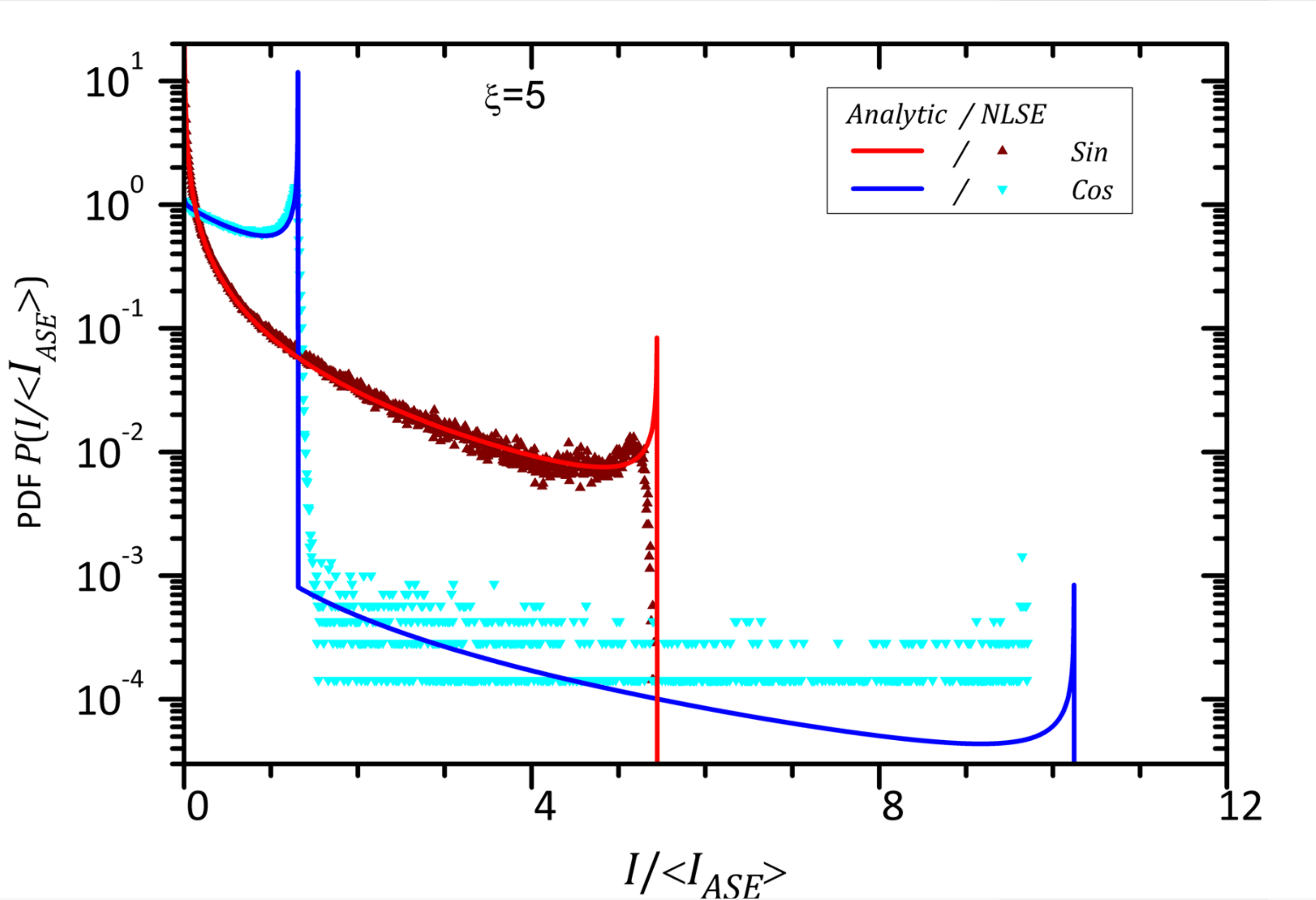}
\caption{
\label{fig:p_s_adv}
Comparison of $p_{s,5}$ and $p_{c,5}$ ($\xi=5$) PDFs obtained by utilizing Eq.~(\ref{eq:composition PDFs deriv}) and (\ref{eq:composition PDFs deriv 2}) (solid curves), and those obtained by computing the histogram of $I_s(t)$ and $I_c(t)$ from NLSE simulations (dotted curves).}
\end{figure}

Hence starting with the same time profile but with the energies mentioned above as initial conditions $I_1(0,t')$ and $I_2(0,t')$, the intensities $I_1(l,t')$ and $I_2(l,t')$ of each arm after propagating through the nonlinear elements (fibers) of length $l$ are obtained by solving the nonlinear Schrödinger equation \cite{AgrawalNLFO, fiberdesk} for the field envelope $A_i(z,t')=\sqrt{I_i(z,t')}\exp(\imath\varphi_i(z,t'))$ ($i=$1, 2), reading
\begin{eqnarray}
\label{eq: NLSE_FD}
    \frac{\partial A(z,t')}{\partial z} =  \sum_{m\geq1}\beta_m \frac{\imath^{m+1}}{m!}\frac{\partial^m}{\partial t'^m}A(z,t')\nonumber \\
    +\imath\gamma\frac{\partial}{\partial t}\left|A(z,t')\right|^2 A(z,t').
\end{eqnarray}
where $z$ is the propagation distance along the fiber, $t'-t-z/v_g$ is the time in the moving frame, $\beta_m$ is the $m$-th order coefficient in the Taylor expansion of the chromatic dispersion $\beta(\omega)$ evaluated at the central angular frequency of the laser $\omega_0=2\pi c/\lambda$, and $\gamma=n_2\omega_0/c\mathcal{A}_\text{eff}$ is the effective nonlinear coefficient due to instantaneous Kerr effect that depends on the nonlinear index $n_2$ of the core material, and the effective modal area $\mathcal{A}_\text{eff}$.

The parameters used in the simulations are a fiber length $l=0.1$~m (with 256 steps), a central laser wavelength $\lambda=1030$~nm. The fiber model correspond to a highly nonlinear air/silica photonic crystal fiber. Silica is characterized by a nonlinear index $n_2=3.2\cdot 10^{-20}$~m${}^2$.W${}^{-1}$. 
For a fiber with a mode field diameter (MFD) of 5~µm ($\mathcal{A}_\text{eff}=\pi\text{MFD}^2/4$) the zero dispersion wavelength (ZDW) is $\lambda_\text{ZDW}\sim \text{1064--1070 nm}$. At the operating wavelength $\lambda=1030$~nm, the dispersion is dominated by the second order term in the Taylor expansion, characterized by a coefficient $\beta_2=6\cdot10^{-3}$~ps${}^{2}$.m${}^{-1}$. 

With the conditions chosen here the intensities may reach few to several tens of GW.cm${}^{-2}$. Using short length $l=0.1$~m limits the influence of the dispersion due to $\beta_2$ in the temporal domain. The intensity profile is nearly unchanged along the propagation. The Kerr effect only affects the temporal phases $\varphi_i(z,t')$ ($i=$1, 2) via self phase modulation (SPM), and $\varphi_i(z,t')\simeq 2\pi n_2 l I_i(z,t')/\lambda$. Here the phases are calculated exactly.

From the NLSE simulated intensities $I_{1,2}(t)$ and phase $\varphi_{1,2}(t)$, the two Sin and Cos MZ outputs intensities $I_c(l,t')$ and $I_s(l,t')$ are respectively given by
\begin{eqnarray}
\label{eq: True Sin}
    I_{c,s}(l,t') &=& \frac{1}{2}\bigg[I_1(l,t')+I_2(l,t')\pm 2\sqrt{I_1(l,t')I_2(l,t')}\times\nonumber \\&{}&\cos\left(\varphi_2(l,t')-\varphi_1(l,t')\right)\bigg],
\end{eqnarray}
where the $+$ sign stands for $I_c$, and the $-$ sign for $I_s$.

The PDFs shown in Fig.~\ref{fig:p_s_adv} are obtained by computing the histograms (frequency count) of $I_s/\langle I_s\rangle$ and $I_c/\langle I_c\rangle$ in the range $[0, 20]$ with 3000 binning intervals. As indicated here, in order to underline the position and the role of the poles $\{\mathcal{I}_k\}$ in the PDF, the output intensities have been normalized with respect to the incident ASE average intensity $\langle I_\text{ASE}\rangle$, contrary to Fig.~\ref{fig:statSources} where the normalization with respect to $\langle I_s\rangle$ and $\langle I_c\rangle$ was intended to highlight the influence of nonlinear filtering on the peak-to-background ratio.

As observed in Fig.\ref{fig:p_s_adv} the agreement between the PDF retrieved form NLSE simulations and the analytical formulae is excellent. Not only the overall variation is well reproduced including the step position associated to the $\{\mathcal{I}_k\}$ poles, but also the absolute and relative heights of the various plateaus are remarkably accurate.

In spite of the huge number of points ($2^{20}$), the statistics is limited for the rarest events. This explains the fuzzy plateau around the density probability of $10^{-4}$ on the Cos curve corresponding to counts close to unity (from 0 to 10 counts for $I/\langle I_\text{ASE}\rangle\gtrsim 1.5$).

The sharp cusps and steps obtained analytically are smoothed in the NLSE simulation. This discrepancy finds its origin in the interplay between the Kerr nonlinearity and the dispersion which is neglected in the analytical derivation above. Indeed, the dispersion tends to spread the peaks in time, and this effect is reinforced by the SPM which locally tends to broaden the spectrum, thus increasing the effect of dispersion. The intensity redistribution in time is consequently smoothing the abrupt transition around each pole responsible for the peak selection.

\section{Higher-order correlations functions $g^{(n)}(\tau)$ of NLF-ASE}
\label{outline:g2gn}

\begin{figure}[b!]
\includegraphics[width=\columnwidth]{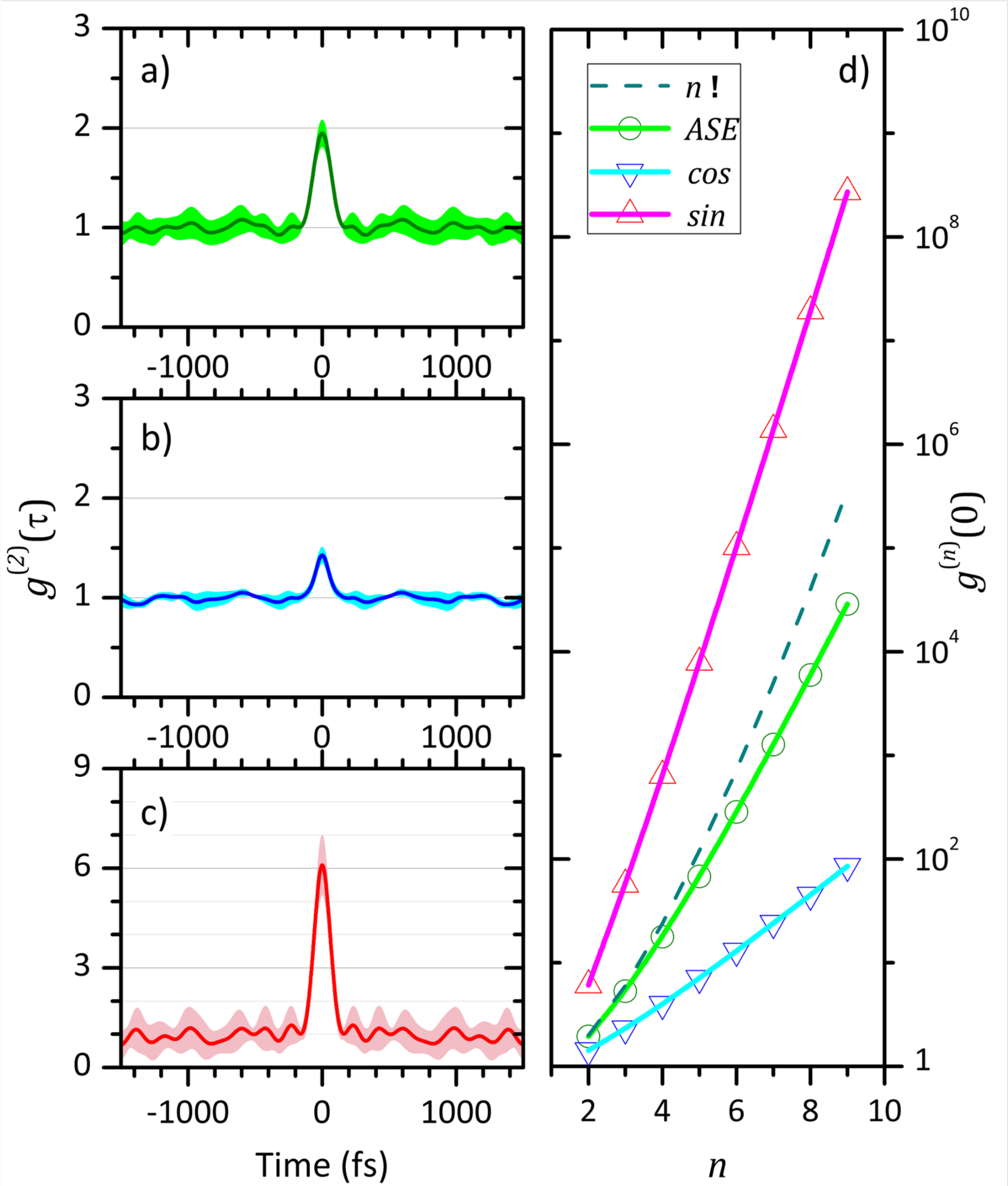}
\caption{
\label{fig:g2gn}
Second-order correlation function $g^{(2)}(\tau)$ for the a) ASE, b) Cos, c) Sin signals. The solid line is the averaged $\langle g^{(2)}(\tau)\rangle_{N_p}$, the shaded ribbon illustrates the standard deviation over $N_p$ independent pulses. (d) Higher-order correlation $g^{(n)}(0)$ as a function of the order $n$ for ASE (open circles $\ocircle$), Sin (upward triangles $\vartriangle$), and Cos (downward triangles $\triangledown$). The dashed line correspond to the theoretical $n$-factorial for incoherent light.}
\end{figure}

As discussed in Section \ref{outline:nlMZ simulation}, $n$th-order correlation functions play a key role in the rate of $n$th-order instantaneous nonlinear processes. The effect of the NLMZ on the $n$th-order correlation functions is illustrated for $n=2$ in Fig.~\ref{fig:g2gn}(a), (b) and (c) that depicts $g^{(2)}(\tau)$ (normalized to 1 for $\tau\rightarrow\pm\infty$) for the ASE, Sin- and Cos-like output respectively. The second order function
\begin{equation}
\label{eq:Def_g_2}
g^{(2)}(\tau)=\left<I(t)\cdot I(t+\tau)\right>_t/\left<I(t)\right>_t^2
\end{equation}
corresponds the autocorrelation that can easily be measured experimentally.

ASE being an incoherent light, it follows the Bose-Einstein distribution \cite{mandel1959fluctuations,pietralunga2003photon}, and $g^{(2)}_\text{ASE}(0)=2$ \cite{Loudon1974,dechatellus2009coherence} as shown in Fig.~\ref{fig:g2gn}(a). Here, $g^{(2)}_\text{Cos}(\tau)$ (Fig.~\ref{fig:g2gn}(b)) and $g^{(2)}_\text{Sin}(\tau)$ (Fig.~\ref{fig:g2gn}(c)) are evaluated for $\xi=5$.
Since we consider $N_p$ independent pulses,
ensemble averaging is taken into account, and the respective average function $\left<g^{(2)}(\tau)\right>_{N_p}$ are plotted, along with their r.m.s fluctuation calculated for $N_p$ individual traces have been scaled up for visibility. Despite the modest value of $n=2$, the effect of nonlinear filtering is already clearly visible, since $\langle g^{(2)}_\text{Sin}(0)\rangle_{N_p}\simeq 3\cdot\langle g^{(2)}_\text{ASE}(0)\rangle_{N_p}$
and $\langle g^{(2)}_\text{Sin}(0)\rangle_{N_p}\simeq 4\cdot\langle g^{(2)}_\text{Cos}(0)\rangle_{N_p}$.

As $g^{(n)}(0) \propto \left<I^n\right>$, higher-order $n$-th order nonlinearity will not only exacerbate the difference of $g^{(n)}(0)$ up to $\gtrsim 6$ orders of magnitude for $n=9$ between Sin, ASE, and Cos  as seen in Fig.~\ref{fig:g2gn}(d), but also the sample-to-sample fluctuations of $g^{(n)}(\tau)$. Eq.~(\ref{eq:Def_g_2}) can be extended to $n$-th order by introducing 
\begin{equation}
\label{eq:Def_g_n_Reduced}
g^{(n)}(\tau)=\left<I^{n-1}(t)\cdot I(t+\tau)\right>_t/\left<I(t)\right>_t^n,
\end{equation}
i.e. the reduced $n$-th order correlation function of the random intensity $I(t)$. Compared to the multivariate degree of $n$-th coherence
\begin{equation}
\label{eq:Def_g_n_Full}
g^{(n)}(\tau_1, \tau_2, \dots, \tau_n)=\frac{\left<I(\tau_1)I(\tau_2)\dots I(\tau_n)\right>_t}{\left<I(t)\right>_t^n},
\end{equation}
the reduced function $g^{(n)}(\tau)$ conveys the more intuitive physical idea of a sequential $(n-1)$-photon $+1$-photon $=n$-photon process. 
Contrary to $g^{(2)}(\tau)$, the expression of $g^{(n)}(\tau)$ and its evaluation on an individual random trial $I(t)$ is not symmetric upon $\tau\rightarrow -\tau$ reversal. Yet, upon statistical averaging, $\left<g^{(n)}(\tau)\right>_{N_p}$ quickly converges (for $N_p>50$) towards a $\tau$-symmetric function (not shown). For all three cases, namely ASE, Sin, and Cos, we have for instance checked that $\left<g^{(n)}(\tau)\right>_{N_p=64}$ converges toward $g^{(n)}(\tau)$ for a $N_p$-times larger sample, also proving the convergence of $g^{(n)}(\tau)$ since time- and ensemble-averaging lead to the same result.

However, $g^{(n)}(\tau)$, and consequently $g^{(n)}(0)$, is computed over a finite time interval. As opposed to the theoretically expected $g^{(n)}_{inc}(0)=n!$ resulting from infinite integration, $g^{(n)}_\text{ASE}(0)$ being only computed over a finite interval also corresponding to the use of pulse-picked sources, its values (open circles) depart from the $n$-factorial law (dashed line) as shown in Fig.~\ref{fig:g2gn}(d). Due to the seldom occurrences of intense peaks, their lack of contribution to the $n$-th power increases with $n$ hence $g^{(n)}_\text{ASE}(0)\leq n!$. In spite of this finite-sample artefact that also affects the Sin- and Cos- output, the enhancement/compression effect can still be observed.

\section{Toy pulses}
\label{sec:toy}
In order to support the 
results of Section \ref{outline:silica absorption} on the energy deposition induced by both ASE and NLF-ASE sources into a dielectric material, simple and representative pulse shapes are studied in this appendix.

\begin{figure}[t!]
\centering
\includegraphics[width=1.\linewidth]{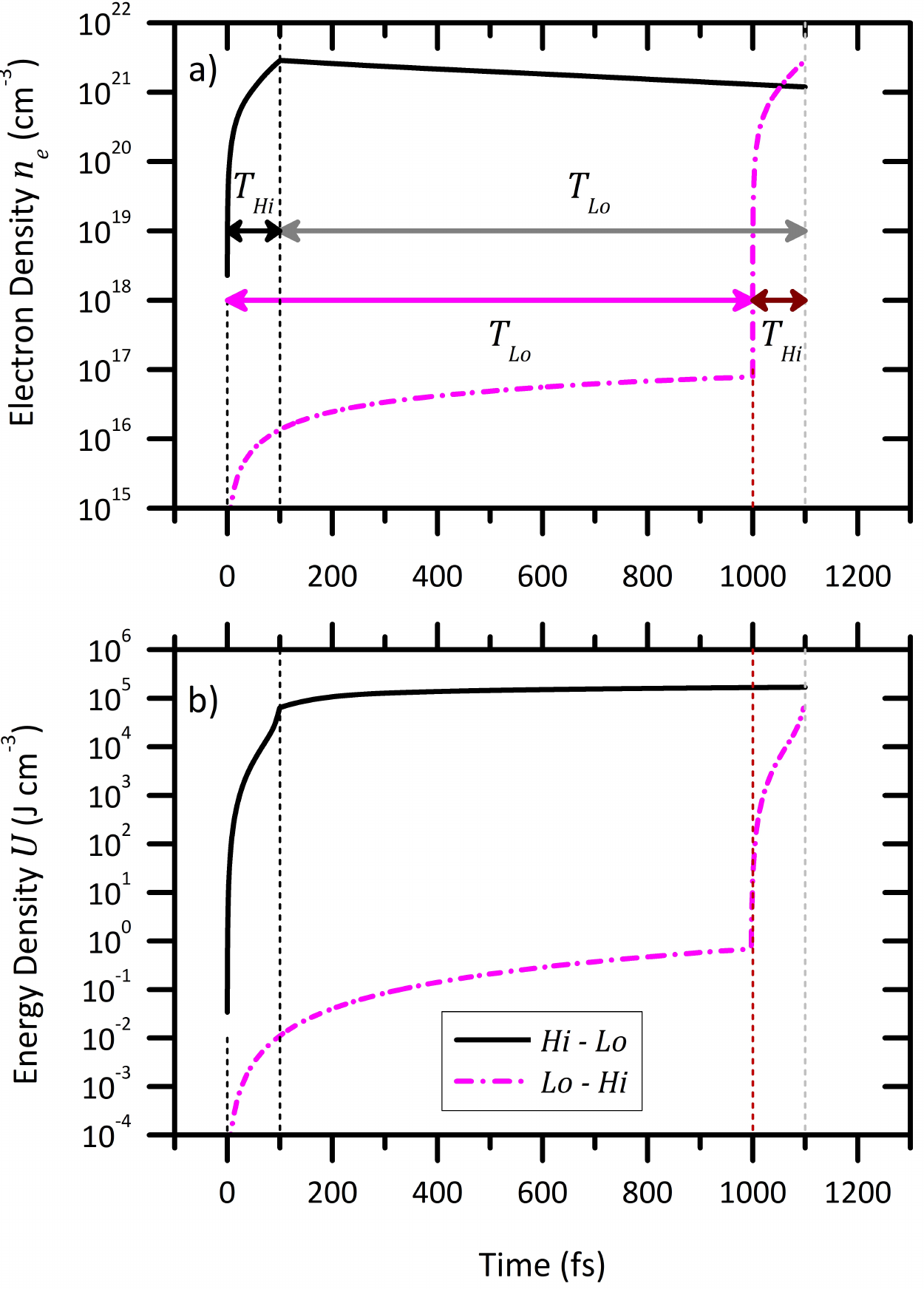}
\caption{Temporal evolution of (a) the free electron density and (b) the energy density induced into the material by a laser pulses shaped with two intensity plateaus. Highest intensity plateau (Hi): $I = 10^{14}$~W/cm$^2$ and $\tau = 100$~fs. Lowest intensity plateau (Lo): $I = 10^{13}$~W/cm$^2$ and $\tau = 1$~ps. The total pulse duration is $1.1$~ps and $\lambda = 1  ~\mu$m. Black and red curves correspond to the plateaus configurations Hi followed by Lo and inversely, respectively.}
\label{fig:ordering}
\end{figure}

The first case aims at exhibiting the influence of pulse shape and time-ordering for a given total fluence. A two-plateau intensity profile is considered where the fluence of each plateau is the same ($10$~J/cm$^2$) but with a factor of $10$ on both intensity and duration; detailed characteristics are provided in Fig. \ref{fig:ordering}. Simulations are performed with the MRE model Sec.(\ref{outline:MRE model}) where the recombination time is set to $500$~fs.

The time-order of plateaus leads to significantly different results. Owing to the nonlinear free electron production with respect to the intensity, the highest-intensity (Hi) plateau leads to the largest electron density and subsequent energy deposition as shown by Fig. \ref{fig:ordering}. When the Hi plateau first irradiates the material, the lowest-intensity (Lo) plateau induces a significant energy deposition due to the larger preexisting electron density.

In the opposite configuration, The Lo plateau does not play any significant role since it produces a negligible electron density and cannot heat the already produced electrons up. In this Lo-Hi configuration, the energy deposition is mainly due to the Hi plateau, resulting in a less efficient interaction compared with the Hi-Lo configuration as seen at $t=1100$~fs in Fig.~\ref{fig:ordering}. These results highlight the importance of time-ordering of sub-pulse intensity structures for the laser energy deposition into the material when cumulative effects are taken into account.

In a second configuration, we consider a $T_\text{burst}=$25 ps long composite pulse structure consisting of a 25~ps flat-top "background" with intensity $I_\text{BG}$ added with a train of 25 short (40~fs) laser pulses. Each sub-pulse has an intensity $I_\text{P}$ in the $10^{14}$~W.cm${}^{-2}$ range and is spaced out of first neighbors by $T_\text{rep}=1$~ps. The contribution of the background and the peaks can be adjusted by varying $I_\text{BG}$ and $I_\text{P}$ while keeping the total fluence constant and equal to 150~J.cm${}^{-2}$.

\begin{figure}[t!]
\centering
\includegraphics[width=1.\linewidth]{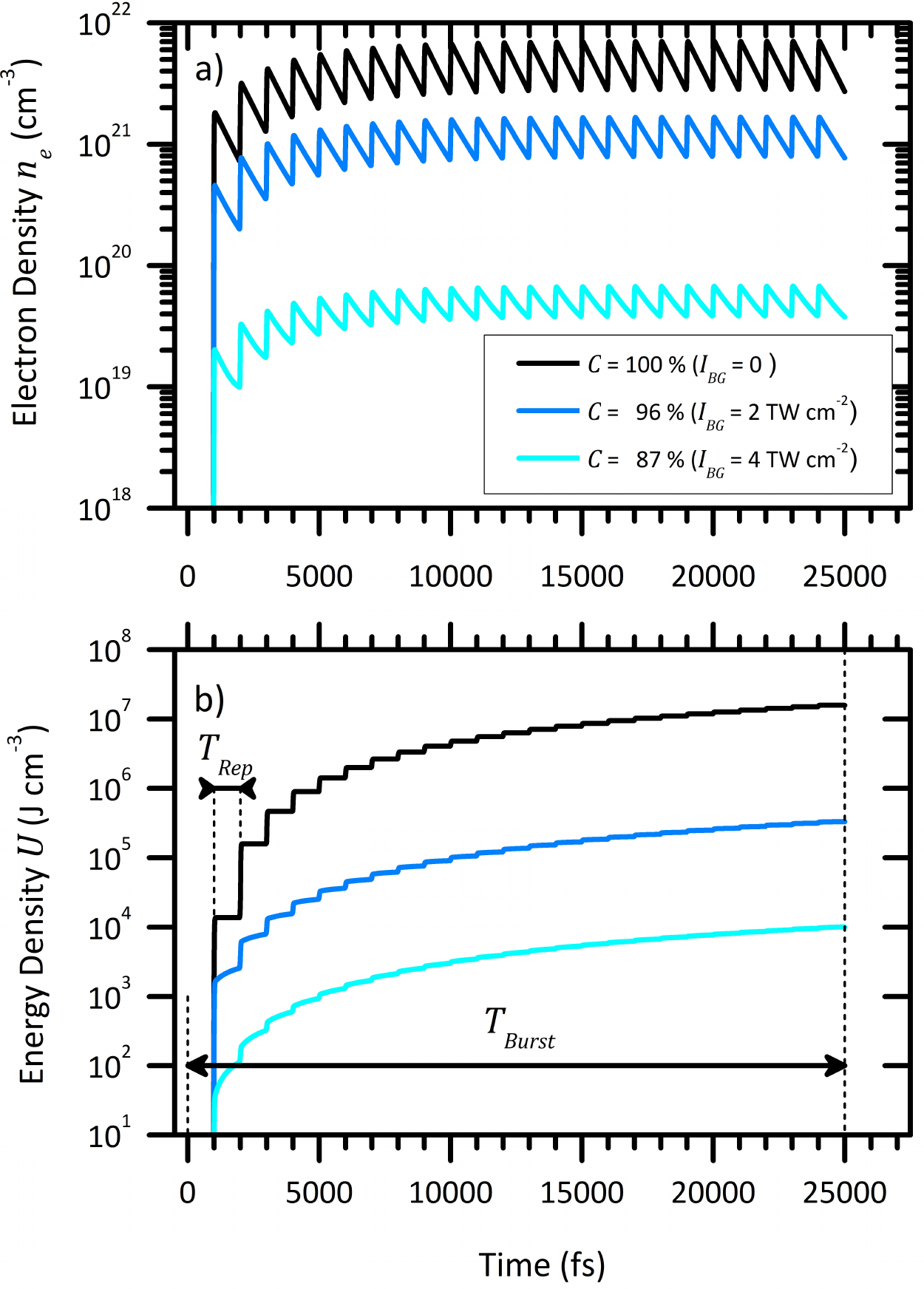}
\caption{Temporal evolution of (a) the free electron density and (b) the energy density induced by a train of laser sub-pulses. The total pulse of duration $25$~ps and fluence $150$~J/cm$^2$ includes $25$ pulses of $40$~fs (FWHM) spaced out by $T_\text{rep} = 1$~ps. Keeping the total fluence constant, a constant intensity background (a few TW/cm$^2$) between sub-pulses has been introduced, resulting in three contrasts. The intensity of each sub-pulse is in the $10^{14}$~W/cm$^2$ range.}
\label{fig:train}
\end{figure}

The motivation for studying such a composite pulse is the following. As discussed, in Sec. \ref{outline:source description} and in Appendix \ref{outline:appendix PDF NLF-ASE}, depending on the output (Sin or Cos) and the value of $\xi$ considered, the PDFs of NLF-ASE sources exhibit a series of "discrete" intensities (see e.g. Fig.~\ref{fig:p_s_adv}). When $\xi=5$, the Sin PDF mostly consists in large peaks ($\gtrsim 10\cdot\left<I\right>$) and nearly no background. For the Cos case fewer large peaks ($\gtrsim 10\cdot\left<I\right>$) are present, and the PDF essentially consists of a continuum of equally probable values between 0 and $2\cdot\left<I\right>$ giving rise to non-zero average background value.

Discarding temporarily the random aspects of NLF-ASE in terms of fluctuations of the peaks intensities (by using a top-hat $I_\text{BG}$ and short pulse of identical $I_\text{P}$) and the delay between consecutive overshoots (by using a periodic pulse train which period $T_\text{rep}=\tau_r$), the composite pulse allows to highlight the effect of the peak-to-background contrast solely.

We thus define the contrast as
\begin{equation}
\label{eq:constrast}
C=\frac{I_\text{P}-I_\text{BG}}{I_\text{P}+I_\text{BG}},
\end{equation}
so that no background ($I_\text{BG}=0$) leads to $C=$100 \%.

Three configurations of contrast $C$ are studied (one with no background and two with $I_\text{BG}=0$ in the TW/cm$^2$ range); detailed characteristics are provided in Fig. \ref{fig:train}. The electron recombination time is set to $\tau_r = 1$~ps, allowing a significant relaxation between sub-pulses (as mentioned before this timescale is consistent with an average time between main intensity peaks of the NLF-ASE source).

As shown by Fig.~\ref{fig:train}(a), regardless of the contrast $C$, the free electron density evolves as steps corresponding to successive significant production events by each sub-pulse followed by the recombination. The intensity background does not significantly affect the recombination dynamics. The evolution of the energy density mimics the electron density as expected. However the intensity background contributes to the evolution of the energy density between sub-pulses, in particular during the first ps as shown by Fig. \ref{fig:train}(b). The lower the contrast, the larger this contribution. Overall the highest contrast leads to the largest free electron density and subsequent energy deposition owing to the non linear feature of the present interaction again.

Within the present laser features, despite long periods of linear laser heating, it appears that the most efficient approach for a significant energy deposition is to produce a large amount of free electrons by intense sub-pulses, the latter being able to further heat electrons. Of course this result does no longer stand when the contrast further decreases because the peak intensity becomes too low. In the extreme case of a zero contrast, i.e. a $25$~ps flat-top pulse with $150$~J/cm$^2$, the produced free electron density is of the order of $10^{16}$~cm$^{-3}$, the energy deposition is $\sim 4$~J/cm$^2$, i.e. many orders of magnitude smaller than values induced by the laser burst.

\bibliographystyle{spiebib}


\end{document}